\documentclass[11pt]{article}
\usepackage[utf8]{inputenc}
\usepackage{mathpazo}
\usepackage{leftindex}
\usepackage{bm}
\usepackage{aurical}
\usepackage[toc,page]{appendix}
\usepackage{latexsym,amsfonts,amsmath,amssymb,mathrsfs,bbold,mathtools,esint,amsthm,mathtools}
\usepackage{dsfont}
\usepackage{mathtools}
\usepackage{float}
\usepackage{graphicx}
\usepackage{cite}
\usepackage{hyperref}
\usepackage{setspace}
\usepackage{color}
\usepackage{slashed}
\usepackage{cleveref}
\numberwithin{equation}{section}
\usepackage[affil-it]{authblk}

\usepackage{float}
\usepackage{indentfirst}
\usepackage{soul}
\usepackage{booktabs}
\usepackage{eso-pic,graphicx}
\usepackage{nicefrac}
\usepackage{epsfig}
\newcommand{\etal}{\textit{et al.}}
\newcommand{\str}{\mathrm{Str}}
\newcommand{\s}{\mathrm{s}}
\newcommand{\co}{\mathrm{c}}
\newcommand{\e}{\mathrm{e}}
\newcommand{\jsn}{\mathrm{sn}}
\newcommand{\jcn}{\mathrm{cn}}
\newcommand{\mi}{\mathrm{i}\,}
\usepackage{braket}
\usepackage[a4paper]{geometry}
\usepackage{a4wide}

\usepackage{color,hyperref}
\definecolor{darkblue}{rgb}{0.0,0.0,0.3}
\hypersetup{colorlinks,breaklinks,linkcolor=darkblue,urlcolor=darkblue,anchorcolor=darkblue,citecolor=darkblue}

\title{On extension of the Yang-Baxter equation and the fermionic $R$-operator}
\author[1]{A. Melikyan\footnote{\href{mailto:amelik@gmail.com}{amelik@gmail.com}}}
\affil[1]{Instituto de Física\\
Universidade de Brasília\\
70910-900, Brasília, DF, Brasil}

\usepackage{palatino}

\begin{document}
\maketitle

\begin{abstract}
We consider the fermionic $R$-operator based on Bazhanov-Stroganov's three-parameter elliptic parametrization of the free fermion model, and find the most general solution of the related tetrahedral Zamolodchikov algebra in the trigonometric limit for an arbitrary set of parameters. This allows to construct an extension of the $R$-operator and the corresponding Yang-Baxter equation, which are of the difference type in one of the spectral parameters.
\end{abstract}


\section{Introduction}
In \cite{Melikyan2020} we considered an elliptic parametrization of the free fermion model on the lattice \cite{Bazhanov:1984iw,Bazhanov:1984ji,Bazhanov:1984jg}, with the $R$-matrices being of the difference type in one of the spectral parameters, i.e., the dependence on the spectral parameters $u_{j}$ is through the differences $\Delta u_{jk}:=u_{j}-u_{k}$, and found a particular solution of the tetrahedral Zamolodchikov algebra. The original motivation came from investigating the integrable properties of the string theory on $AdS_5 \times S^5$ (see \cite{Beisert:2010jr} for a review), and, in particular, its relation to the one-dimensional Hubbard model \cite{Essler:2005bk}, the $R$-matrix of which  has been proposed to be related to the $S$-matrix of the string theory via $AdS/CFT$ correspondence (see \cite{Mitev2017,Beisert:2006qh,Rej2006} and the references therein). It has been long known that the $R$-matrix of the one-dimensional Hubbard model can be obtained, as shown by Shastry \cite{Shastry:1986zz,Shastry1988}, by pairing two $R$-matrices corresponding to the free fermion model. The construction, which is based on the use of the so-called decorated Yang-Baxter equation ($DYBE$), in addition to the standard Yang-Baxter equation ($YBE$), leads to an $R$-matrix which is not of  the difference type. 

There is another interesting integrable fermionic model which appears in  the $su(1 \vert 1)$  subsector of the string theory  on $AdS_5 \times S^5$, which is described by a $(1+1)$-relativistic fermion model \cite{Arutyunov:2004yx}.\footnote{ The Lagrangian of the model has the following explicit form:
\begin{align}
	 \mathscr{L}&= \mi \bar{\psi} \gamma_{\mu} \partial^{\mu} \psi \: - m \bar{\psi} \psi + \frac{g_2}{4m} \epsilon^{\alpha \beta} \left( \bar{\psi}
	\partial_{\alpha} \psi \; \bar{\psi}\: \gamma^3 
	\partial_{\beta} \psi -
	\partial_{\alpha}\bar{\psi} \psi \; 
	\partial_{\beta} \bar{\psi}\: \gamma^3 \psi \right)-  \frac{g_3}{16m} \epsilon^{\alpha \beta} \left(\bar{\psi}\psi\right)^2 
	\partial_{\alpha}\bar{\psi}\:\gamma^3
	\partial_{\beta}\psi. \label{intro:aaf_lag}
\end{align}}
It can be shown to be a completely  integrable classically, and its quantum integrability properties has been investigated from various points of view \cite{Melikyan:2011uf,Melikyan:2012kj,Melikyan:2014yma,Melikyan2019ac,Melikyan:2016gkd,Melikyan:2014mfa}. Nevertheless, due to the non-ultralocal nature of the algebra of Lax operators,\footnote{This remains the case if one sets $g_{2}=g_{3}=0$, i.e., the non-ultralocal nature is also present for the free fermion model. The explicit forms of the Lax operators and the corresponding algebra can be found in \cite{Melikyan:2012kj,Melikyan:2014yma}.} the quantization of this model by standard methods of integrable systems remains an open problem, and, as a consequence, the lattice version of the model is not known. One can, however, start with the lattice formulation of a model - in particular, given the same non-ultralocal nature for the free fermion and the model in \eqref{intro:aaf_lag} - with the free fermion model on the lattice and trace the appearance of the non-ultralocal terms in the continuous limit.\footnote{A partial progress towards this program was reported in \cite{Melikyan:2020ibw}.} For the interacting theory, the one-dimensional Hubbard model is an obvious candidate, given, as mentioned above, its relation to string theory on $AdS_5 \times S^5$ on one hand, and to the free fermion model, by means Shastry's procedure of pairing two corresponding $R$-matrices, on the other hand (see, however, \cite{Frahm:1990ab,Links:2001ab}). Thus, it would be quite advantageous to start with the $R$-matrix for the free fermion model of the difference type, and investigate the consequences when coupling the spins, as, for example, in the case of the one-dimensional Hubbard model, which would in principle allow obtaining the lattice formulations of interacting continuous $(1+1)$-relativistic fermion models such as the model in \eqref{intro:aaf_lag}.

To this end, we consider the fermionic $R$-operator formalism  due to Umeno $\etal$ \cite{Umeno1998b,Umeno1998,Umeno2000}, and the three-parameter elliptic parametrization of the free fermion model due to Bazhanov and Stroganov \cite{Bazhanov:1984iw,Bazhanov:1984ji,Bazhanov:1984jg}, which allow a formulation of both  Yang-Baxter and decorated  Yang-Baxter relations in the form where the dependence of  $R$-matrix is of the difference type with respect to one of the spectral parameters. We give the main formulas and results of this construction in section \ref{bs}. Next, in section \ref{tza}, we give the most general solution of the tetrahedral Zamolodchikov algebra \cite{Zamolodchikov1981,Korepanov1993,Korepanov2013,Korepanov1994c} in the trigonometric limit for an arbitrary set of parameters. Using these results, in section \ref{eybe}, we construct an extension of the original fermionic $R$-operator and obtain a family of the corresponding Yang-Baxter equations, each being of the difference type in one of the spectral parameters. In appendices we collect many explicit formulas and useful expressions: In Appendix \ref{app-id} we give some useful identities used in the text, in Appendix \ref{app-tza} we list the explicit form of the equations arising in the tetrahedral Zamolodchikov algebra for the full elliptic case, and, finally, in Appendix \ref{app-coef} we present our solution to the tetrahedral Zamolodchikov algebra in the trigonometric limit for an arbitrary set of spectral parameters.

\section{ Three-parameter elliptic parametrization of the free fermion model}
\label{bs}

In \cite{Bazhanov:1984iw,Bazhanov:1984ji,Bazhanov:1984jg} (see also \cite{Bazhanov2012}) an interesting general solution to an inhomogeneous eight-vertex model was obtained, with the  $R$-matrix of the form: 
\begin{align}
    \hat{R}=\begin{pmatrix}
a & 0 & 0 & d\\
0 & b & c & 0 \\
0 & c^{\prime} & b^{\prime} & 0 \\
d^{\prime} & 0 & 0 & a \label{bs:R_matrix_orig}
\end{pmatrix},
\end{align}
satisfying the free fermion condition \cite{Fanwu723}:
\begin{align}
    a a^{\prime} + b b^{\prime} - c c^{\prime} - d d^{\prime}=0 \label{bs:free_fermion_condition}
\end{align}
The  coefficients in \eqref{bs:R_matrix_orig} are parametrized in terms of by three parameters: $\left\{\zeta_{0}\equiv u,\zeta_{1},\zeta_{2}\right\}$, and have the form:
\begin{align}
    &a(u;\zeta_{1},\zeta_{2}) =\rho \left[ 1-\e(u)\e(\zeta_{1})\e(\zeta_{2}) \right]; \quad a^{\prime}(u;\zeta_{1},\zeta_{2}) =\rho \left[ \e(u)-\e(\zeta_{1})\e(\zeta_{2}) \right] \label{bs:a_a_prime} \\
    &b(u;\zeta_{1},\zeta_{2}) =\rho \left[ \e(\zeta_{1})-\e(u)\e(\zeta_{2}) \right]; \quad b^{\prime}(u;\zeta_{1},\zeta_{2}) =\rho \left[ \e(\zeta_{2})-\e(u)\e(\zeta_{1}) \right] \label{bs:b_b_prime}\\
    &c(u;\zeta_{1},\zeta_{2})=c^{\prime}(u;\zeta_{1},\zeta_{2}) =\rho \: \jsn^{-1}\left(\frac{u}{2}\right)\left[ 1-\e(u)\right]\left[\e(\zeta_{1})e(\zeta_{2})\jsn(\zeta_{1})\jsn(\zeta_{2}) \right]^{1/2}, \label{bs:c_c_prime}\\
    &d(u;\zeta_{1},\zeta_{2})=d^{\prime}(u;\zeta_{1},\zeta_{2}) =- \mi k \rho \: \jsn \left(\frac{u}{2} \right)\left[ 1+\e(u)\right]\left[\e(\zeta_{1})e(\zeta_{2})\jsn(\zeta_{1})\jsn(\zeta_{2}) \right]^{1/2} \label{bs:d_d_prime}.
\end{align}
Here the functions $\jsn(x)$ and $\jcn(x)$ are the Jacobi elliptic functions of modulus $k$ \cite{whittaker_watson_1996}, $\e(x)$ is the elliptic exponential $\e(x):=\jcn(x) + \mi \jsn(x)$, and $\rho$ is an arbitrary factor.
The $R$-matrix in \eqref{bs:R_matrix_orig} satisfies the Yang-Baxter equation \cite{Bazhanov:1984iw}:
\begin{align}
   \hat{R}_{12}(u_{12};\zeta_{1},\zeta_{2})\hat{R}_{13}(u_{13};\zeta_{1},\zeta_{3})\hat{R}_{23}(u_{23};\zeta_{2},\zeta_{3})=\hat{R}_{23}(\eta_{23};\zeta_{2},\zeta_{3})\hat{R}_{13}(u_{13};\zeta_{1},\zeta_{3})\hat{R}_{12}(u_{12};\zeta_{1},\zeta_{2}),\label{bs:YBE}
\end{align}
where we have used the shorthand notation $u_{jk} \equiv u_{j}-u_{k}$. 

Following \cite{Umeno1998b,Umeno1998}, we next introduce the fermionic $R$-operator corresponding to the $R$-matrix \eqref{bs:R_matrix_orig}, which is more convenient to use from the beginning in order, e.g., to make a connection with the Lax operators for the relativistic $(1+1)$ free fermion model or the model defined by \eqref{intro:aaf_lag}, which are written in terms of fermionic variables. To this end, one has to apply the Jordan-Wigner transformation (see \cite{Essler:2005bk} for an extensive treatment) to the above $\hat{R}$-matrix as well as the the Yang-Baxter equation \eqref{bs:YBE}. The essential technical steps are explained in details in \cite{Umeno1998b,Umeno1998} and are omitted here. 

The final result, when applied to the $R$-matrix \eqref{bs:R_matrix_orig} is as follows: The associated fermionic $R$-operator, written in terms of the fermionic variables $(c^{\dagger}_{k},c_{k})$, satisfying the usual anti-commutation relations: $\{c^{\dagger}_{j},c_{k}\}=\delta_{jk}$, takes the from:
\begin{align}
    R_{jk}(u;\zeta_{j},\zeta_{k})&=a(u;\zeta_{j},\zeta_{k})\left[-n_{j} n_{k} \right] +a^{\prime}(u;\zeta_{j},\zeta_{k})\left[(1-n_{j})(1- n_{k}) \right]+b(u;\zeta_{j},\zeta_{k})\left[n_{j}(1- n_{k}) \right] \nonumber\\
    &+b^{\prime}(u;\zeta_{j},\zeta_{k})\left[n_{k}(1- n_{j})\right] +c(u;\zeta_{j},\zeta_{k})\left[\Delta_{jk}+\Delta_{kj} \right] + d(u;\zeta_{j},\zeta_{k})\left[-\tilde{\Delta}^{(\dagger)}_{jk}-\tilde{\Delta}_{jk}\right].\label{bs:fermionic_R}
\end{align}
Here we have denoted: 
\begin{align}\label{bs:notations}
  n_{k}&:=c^{\dagger}_{k}c_{k},\quad \Delta_{jk}:=c^{\dagger}_{j}c_{k},\quad
\tilde{\Delta}^{(\dagger)}_{jk}=c^{\dagger}_{j}c^{\dagger}_{k}; \quad \tilde{\Delta}_{jk}=c_{j}c_{k}.
\end{align}
Furthermore, it can be shown that the fermionic $R$-operator \eqref{bs:fermionic_R} satisfies the Yang-Baxter equation:
\begin{align}
    R_{12}(u_{12};\zeta_{1},\zeta_{2})R_{13}(u_{13};\zeta_{1},\zeta_{3})R_{23}(u_{23};\zeta_{2},\zeta_{3})=R_{23}(u_{23};\zeta_{2},\zeta_{3})R_{13}(u_{13};\zeta_{1},\zeta_{3})R_{12}(u_{12};\zeta_{1},\zeta_{2}),\label{bs:YBE_fermionic}
\end{align}
The corresponding fermionic monodromy operator is defined in the usual manner:
\begin{align}
    T_{a}(u;\zeta_{a},\{\zeta_{j}\})=R_{aN}(u;\zeta_{a},\zeta_{N})R_{a,N-1}(u;\zeta_{a},\zeta_{N-1})\cdot\ldots \cdot R_{a1}(u;\zeta_{a},\zeta_{1}),\label{bs:monodromy}
\end{align}
and satisfies the $RTT=TTR$ relation:
\begin{align}
     R_{ab}(u_{a}-u_{b};\zeta_{a},\zeta_{b}) T_{a}(u_{a};\zeta_{a},\{\zeta_{j}\}) T_{b}(u_{b};\zeta_{b},\{\zeta_{j}\})=T_{b}(u_{b};\zeta_{b},\{\zeta_{j}\})T_{a}(u_{a};\zeta_{a},\{\zeta_{j}\})R_{ab}(u_{a}-u_{b};\zeta_{a},\zeta_{b}). \label{bs:rtt_ttr}
\end{align}
Defining also the transfer operator:
\begin{align}
    \tau(u;\zeta_{a},\{\zeta_{j}\})=\str_{a}\left[ T_{a}(u;\zeta_{a},\{\zeta_{j}\})\right], \label{bs:tau}
\end{align}
where the supertrace $\str_{a}\left[F\right]$ over the auxiliary Fock space $a$ is defined by \cite{Umeno1998b}:
\begin{align}\label{bs:supertrace_spinless}
    \str_{a}\left[F\right]&:={}_a\bra{0}F \ket{0}_{a} -{}_a\bra{1} F \ket{1}_{a}\\
    c_{a} \ket{0}_{a} &=0,\quad \ket{1}_{a}:= c^{\dagger}_{a} \ket{0}_{a}.
\end{align}
one obtains the commuting quantities:
\begin{align}\label{bs:tau_commute}
    \left[\tau(u;\zeta_{a},\{\zeta_{j}\}),\tau(v;\zeta_{b},\{\zeta_{j}\}) \right]=0.
\end{align}
As an example, we apply the fermionic Yang-Baxter relation \eqref{bs:YBE_fermionic} for the case  of equal parameters $\zeta_{i} \equiv \zeta$, and compute the Hamiltonian via the standard formula \cite{Essler:2005bk}:
\begin{align}
    \hat{\mathcal{H}}=\tau^{-1}(0;\zeta)\frac{d}{du}\tau(u;\zeta)\vert_{u=0}. \label{bs:general_formula}
\end{align}
Using the explicit form of the coefficients \eqref{bs:a_a_prime}-\eqref{bs:d_d_prime} it is easy to check the relations:
\begin{align}
    &R_{jk}(0;\zeta)=\beta \mathcal{P}_{jk}\label{bs:regularity}\\
   &\mathcal{P}_{jk}R^{\prime}(0;\zeta)=a^{\prime}_{0}(\zeta)+\left\{c_{0}(\zeta)-a^{\prime}_{0}(\zeta)\right\}\left[n_{j}+n_{k}\right]+b_{0}(\zeta)\left[\Delta_{jk}+\Delta_{kj} \right]+ d_{0}(\zeta)\left[\tilde{\Delta}^{(\dagger)}_{jk}-\tilde{\Delta}_{jk}\right]\label{bs:fermionic_R_der}\\
&\tau(0;\zeta)=\beta^{N}\mathcal{P}_{12}\mathcal{P}_{23}\cdot \ldots \cdot \mathcal{P}_{N,N-1},\label{bs:tau_0}
\end{align}
where we denoted $R_{jk}(0;\zeta)=R_{jk}(0;\zeta,\zeta)$, $\beta=(-2 \mi \rho)\e(\zeta) \jsn(\zeta)$,  $\mathcal{P}_{jk}:=1-n_{j}-n_{k}+\Delta_{jk}+\Delta_{kj}$ is the permutation operator, and $a^{\prime}_{0}(\zeta),c_{0}(\zeta),b_{0}(\zeta),d_{0}(\zeta)$ in the right-hand side of \eqref{bs:fermionic_R_der} stand for the derivatives of  the corresponding coefficients \eqref{bs:a_a_prime}-\eqref{bs:d_d_prime} at $u=0$. Using \eqref{bs:general_formula}-\eqref{bs:tau_0}, one immediately obtains the Hamiltonian for the fermionic $XY$ model in an external field:\footnote{Note, that the parameter $\zeta$ controls the external field in \eqref{bs:XY_Hamiltonian}.}
\begin{align}
   \hat{\mathcal{H}}^{{(XY)}}= \frac{1}{2\,\jsn(\zeta)}\sum_{j=1}^{N}\left[\left(\Delta_{j,j+1}+\Delta_{j+1,j}\right)+ k\,\jsn(\zeta) \left(\tilde{\Delta}^{(\dagger)}_{j,j+1}-\tilde{\Delta}_{j+1,j}\right)+2\,\jcn(\zeta)\left(n_{j}-\nicefrac{1}{2}\right)\right]. \label{bs:XY_Hamiltonian}
\end{align}

To extend the above construction for the spinless fermions to the models with spin, one introduces two copies of the $R^{(s)}$-operator for each spin $s=\ket{\uparrow},\ket{\downarrow}$. The generalization is straightforward, and the free fermion model with spin $s=\nicefrac{1}{2}$ is obtained from two copies of the fermionic $R$-matrix \eqref{bs:fermionic_R}:
\begin{align}
   \mathcal{R}_{jk}(u_{j}-u_{k};\zeta_{j},\zeta_{k};\hat{\zeta}_{j},\hat{\zeta}_{k}):=R^{(\uparrow)}_{jk}(u_{j}-u_{k};\zeta_{j},\zeta_{k})R^{(\downarrow)}_{jk}(u_{j}-u_{k};\hat{\zeta}_{j},\hat{\zeta}_{k}). \label{bs:fermionic_R_spin}
\end{align}
The fermionic operator $\mathcal{R}_{jk}(u;\zeta_{j},\zeta_{k})$ in \eqref{bs:fermionic_R_spin} satisfies the same Yang-Baxter equation \eqref{bs:YBE_fermionic}, and the corresponding monodromy operator and commuting quantities can be constructed in the same manner as above. Starting from $\mathcal{R}_{jk}(u;\zeta_{j},\zeta_{k};\hat{\zeta}_{j},\hat{\zeta}_{k})$, corresponding to the case of equal parameters $\zeta_{i} \equiv \zeta,\hat{\zeta}_{i} \equiv \hat{\zeta} $ considered above, one obtains two copies of non-interacting fermionic $XY$ models in external fields, with the $\mathcal{R}$-matrix \eqref{bs:fermionic_R_spin} being of the difference type in one of the spectral parameters, unlike the case in \cite{Umeno1998b,Umeno1998}.\footnote{In this case, the definition of the supertrace over the auxiliary Fock space $a$ (c.f. \eqref{bs:supertrace_spinless}) is as follows \cite{Umeno1998b}:
\begin{align}\label{bs:supertrace_spin}
    \str_{a}\left[F\right]&:={}_a\bra{0} F \ket{0}_{a} -{}_a\bra{\uparrow} F \ket{\uparrow}_{a} -{}_a\bra{\downarrow} F \ket{\downarrow}_{a}+{}_a\bra{\downarrow \uparrow} F \ket{\uparrow \downarrow}_{a}\\
    c_{a,(s)} \ket{0}_{a} &=0,\quad \bra{\uparrow}_{a}:= c^{\dagger}_{a,\uparrow}\ket{0}_{a},\quad \bra{\downarrow}_{a}:= c^{\dagger}_{a,\downarrow} \ket{0}_{a},\quad \bra{\uparrow \downarrow}_{a}:= c^{\dagger}_{a,\uparrow}c^{\dagger}_{a,\downarrow}\ket{0}_{a}.\nonumber
\end{align}
} The details of obtaining the corresponding Lax pair were given in \cite{Melikyan:2020ibw}, and the problem of taking the continuous limit will be presented in a future publication.

\section{Decorated Yang-Baxter equation and tetrahedral Zamolodchikov algebra}
\label{tza}

We now turn our attention to the interacting case. First, we derive the so-called decorated Yang-Baxter equation \cite{Shastry:1986zz,Shastry1988,Umeno1998b}. In our case, it follows from the relation:\footnote{Changing the sign of the modulus $k \to -k$ only changes the sign of $d(u;\zeta_{1},\zeta_{2})$ in \eqref{bs:d_d_prime}.}
\begin{align}
    (2n_{j,(s)}-1)R^{(s)}_{jl}(u;\zeta_{j},\zeta_{l};k) (2n_{l,(s)}-1)=R^{(s)}_{jl}(u;\zeta_{j}+2 \mathrm{K}(k),\zeta_{l}+2 \mathrm{K}(k);-k);\quad s=\ket{\uparrow},\ket{\downarrow} \label{tza:identity_main}
\end{align}
where we have written the dependence on the modulus $k$ in $R^{(s)}_{jl}(u;\zeta_{j},\zeta_{k};k)$ explicitly, and $\mathrm{K}(k)$ is the complete elliptic integral of the first kind \cite{whittaker_watson_1996}. The above formula can be checked using the formulas given in  Appendix \ref{app-id}, as well as  the explicit expressions for the functions \eqref{bs:a_a_prime}-\eqref{bs:d_d_prime}. Then, the decorated Yang-Baxter equation is a relation that is derived from the Yang-Baxter equation \eqref{bs:YBE_fermionic}, together with the above identity \eqref{tza:identity_main}. It has the form:
\begin{align}
    R^{(s)}_{12}(u_{12};\zeta_{1},\zeta_{2}-2\mathrm{K}(k);k) (2n_{1,(s)}-1) R^{(s)}_{13}(u_{13};\zeta_{1},\zeta_{3}-2\mathrm{K}(k);-k) R^{(s)}_{23}(u_{23};\zeta_{2},\zeta_{3};k) \nonumber \\
    =R^{(s)}_{23}(u_{23};\zeta_{2},\zeta_{3};k)R^{(s)}_{13}(u_{13};\zeta_{1},\zeta_{3}-2\mathrm{K}(k);-k)(2n_{1,(s)}-1)R^{(s)}_{12}(u_{12};\zeta_{1},\zeta_{2}-2\mathrm{K}(k);k) \label{tza:DYBE}
\end{align}
Unlike the previously considered cases \cite{Essler:2005bk,Shastry:1986zz,Shastry1988,Umeno1998b}, the decorated Yang-Baxter relation \eqref{tza:DYBE} depends on the differences of the spectral parameters  $u_{jk} \equiv u_{j}-u_{k}$, taking an asymmetrical form with respect to the other parameters $\zeta_{i}$. Using \eqref{tza:identity_main} and taking the product of two copies of \eqref{tza:DYBE} for $s=\ket{\uparrow},\ket{\downarrow}$ one can readily arrive at a similar identity for $\mathcal{R}_{jk}(u;\zeta_{j},\zeta_{k})$ defined in \eqref{bs:fermionic_R_spin}.
In what follows, it is convenient to introduce the following notations:
\begin{align}
    \mathrm{L}^{0,(s)}_{jk} &=R^{(s)}_{jk}(u_{jk};\zeta_{j},\zeta_{k};k) \label{tza:L0}\\
    \mathrm{L}^{1,(s)}_{jk} &=R^{(s)}_{jk}(u_{jk};\zeta_{j},\zeta_{k}-2\mathrm{K}(k);-k)(2n_{j,s}-1), \label{tza:L1}
\end{align}
The explicit form of the $ \mathrm{L}^{1,(s)}_{jk}$-operator is given in \eqref{app:fermionic_R_id1}, Appendix \ref{app-id}. With these notation, the Yang-Baxter and decorated Yang-Baxter equations take the form:
\begin{align}
    \mathrm{L}^{0,(s)}_{12} \mathrm{L}^{0,(s)}_{13} \mathrm{L}^{0,(s)}_{23} &=  \mathrm{L}^{0,(s)}_{23} \mathrm{L}^{0,(s)}_{13} \mathrm{L}^{0,(s)}_{12}, \label{tza:L000} \\
     \mathrm{L}^{0,(s)}_{12} \mathrm{L}^{1,(s)}_{13} \mathrm{L}^{1,(s)}_{23} &=  \mathrm{L}^{1,(s)}_{23} \mathrm{L}^{1,(s)}_{13} \mathrm{L}^{0,(s)}_{12}. \label{tza:L011}
\end{align}
The tetrahedral Zamolodchikov algebra is an algebraic expression of the form:
\begin{align}
    \mathrm{L}^{\alpha_{1},(s)}_{12} \mathrm{L}^{\alpha_{2},(s)}_{13} \mathrm{L}^{\alpha_{3},(s)}_{23} =\sum_{\beta_{i}=0,1} S^{\alpha_{1} \alpha_{2} \alpha_{3}}_{\beta_{1}\beta_{2}\beta_{3}}\mathrm{L}^{\beta_{1},(s)}_{23} \mathrm{L}^{\beta_{2},(s)}_{13} \mathrm{L}^{\beta_{3},(s)}_{12}; \quad \alpha_{1,2,3}=\{0,1\},\label{tza:TZA}
\end{align}
generalizing the above two relations \eqref{tza:L000} and \eqref{tza:L011}. It was introduced by Korepanov in \cite{Korepanov:1989tl,Korepanov1993,Korepanov1994c,Korepanov1994,Korepanov2013} to investigate Zamolodchikov's tetrahedron equation, which underlines the symmetry of a three-dimensional integrable model, generalizing the relation of the Yang-Baxter equation to two-dimensional integrable models. As was shown in \cite{Shiroishi1995as,Shiroishi1995vc,Umeno1998b} the tetrahedral Zamolodchikov algebra can be used in order to construct an interacting model of spin $s=\nicefrac{1}{2}$ fermions, and, in particular, to obtain the one-dimensional Hubbard model (see \cite{Essler:2005bk} for a review, and there references therein). Thus, our first goal is to obtain the solutions to tetrahedral algebra \eqref{tza:TZA}.
To obtain the coefficients $S^{\alpha_{1} \alpha_{2} \alpha_{3}}_{\beta_{1}\beta_{2}\beta_{3}}$  in \eqref{tza:TZA} one has to evaluate generic tensor products $\mathrm{L}^{\alpha_{1},(s)}_{12} \mathrm{L}^{\alpha_{1},(s)}_{13} \mathrm{L}^{\alpha_{3},(s)}_{23}$ and $\mathrm{L}^{\beta_{1},(s)}_{23} \mathrm{L}^{\beta_{1},(s)}_{13} \mathrm{L}^{\beta_{3},(s)}_{12}$ that appear in the left and right hand sides of \eqref{tza:TZA}. To this end, we note that both $\mathrm{L}^{0,(s)}_{jk}$ and $\mathrm{L}^{1,(s)}_{jk}$ can be written in the following general form:
\begin{align}
    \Gamma_{jk}(u;\zeta_{j},\zeta_{k}) &=a_{0}(u;\zeta_{j},\zeta_{k})+a_{1}(u;\zeta_{j},\zeta_{k})n_{j} +a_{2}(u;\zeta_{j},\zeta_{k})n_{k}
    +a_{3}(u;\zeta_{j},\zeta_{k})n_{j}n_{k} +c_{1}(u;\zeta_{j},\zeta_{k})\Delta_{jk} \nonumber \\ &+c_{2}(u;\zeta_{j},\zeta_{k})\Delta_{kj} +d_{1}(u;\zeta_{j},\zeta_{k})\tilde{\Delta}^{(\dagger)}_{jk}+d_{2}(u;\zeta_{j},\zeta_{k})\tilde{\Delta}_{jk} ,\label{tza:fermionic_R_generic_form}
\end{align}
One then obtains either $\mathrm{L}^{0,(s)}_{jk}$ or $\mathrm{L}^{1,(s)}_{jk}$ by fixing accordingly the functions $\Bigl(a_{0}(u;\zeta_{j},\zeta_{k}),\ldots,\\
a_{3}(u;\zeta_{j},\zeta_{k}),c_{1,2}(u;\zeta_{j},\zeta_{k}),d_{1,2}(u;\zeta_{j},\zeta_{k})\Bigr)$ in terms of the coefficients of the $R$-operator \eqref{bs:a_a_prime}-\eqref{bs:d_d_prime}. Thus, one only needs to evaluate the general tensor products of the form:
\begin{align}
    \Omega&:=\Gamma_{12}(w;\zeta^{\prime \prime}_{1},\zeta^{\prime \prime}_{2}) \Gamma_{13}(u;\zeta_{1},\zeta_{3}) \Gamma_{23}(v;\zeta^{\prime}_{2},\zeta^{\prime}_{3})\\ \tilde{\Omega}&:=\Gamma_{23}(\hat{v};\hat{\zeta}_{2},\hat{\zeta}_{3}) \Gamma_{13}(\hat{u};\hat{\zeta}^{\prime}_{1},\hat{\zeta}^{\prime}_{3})\Gamma_{12}(\hat{w};\hat{\zeta}^{\prime \prime}_{1},\hat{\zeta}^{\prime \prime}_{2})
\end{align}
to take into account all possible permutations of indices $\alpha_{1,2,3}$ and $\beta_{1,2,3}$ in \eqref{tza:TZA}. 
These expressions are also needed to verify the linear dependency of the $\mathrm{L}^{(0,1),(s)}_{jk}$-operators products. To this end, it is convenient to represent the explicit expressions for $\Omega$, and, similarly, for $\tilde{\Omega}$, as sums of terms corresponding to all possible independent products of operators $\mathcal{O}_{a}=\{\mathbb{1},n_{i},\Delta_{jk},\tilde{\Delta}_{jk},\tilde{\Delta}^{(+)}_{jk}\}$:
\begin{align}
    \Omega &=\sum_{ab} \left[\Gamma_{\mathcal{O}_{a}\mathcal{O}_{b}}\right] \mathcal{O}_{a}\mathcal{O}_{b} = \Gamma^{(0)}_{0} + \left[\Gamma^{(0)}_{n_{1}}\right] n_{1} + \left[\Gamma^{(0)}_{n_{1}n_{2}}\right] n_{1}n_{2} +  \ldots + \left[\Gamma^{(1)}_{\Delta_{12}}\right] \Delta_{12}+ \left[\Gamma^{(1)}_{\tilde{\Delta}_{12}}\right]\tilde{\Delta}_{12}+ \ldots ,\label{tza:Omega}\\
    \tilde{\Omega} &= \sum_{ab} \left[\tilde{\Gamma}_{\mathcal{O}_{a}\mathcal{O}_{b}}\right] \mathcal{O}_{a}\mathcal{O}_{b} = \tilde{\Gamma}^{(0)}_{0} + \left[\tilde{\Gamma}^{(0)}_{n_{1}}\right] n_{1} + \left[\tilde{\Gamma}^{(0)}_{n_{1}n_{2}}\right] n_{1}n_{2} +  \ldots + \left[\tilde{\Gamma}^{(1)}_{\Delta_{12}}\right] \Delta_{12}+ \left[\tilde{\Gamma}^{(1)}_{\tilde{\Delta}_{12}}\right]\tilde{\Delta}_{12}+ \ldots \, ,\label{tza:Omegatilde}
\end{align}
where we use the following notations for the coefficients $\left[{\Gamma}^{A}_{B}\right]$ in the above expansion: the subscript $B$ indicates the operator which multiplies $\left[{\Gamma}^{A}_{B}\right]$, and the superscript $A$ in indicates the presence ($A=1$) or absence $(A=0)$ of $\{\Delta_{jk},\tilde{\Delta}_{jk},\tilde{\Delta}^{(+)}_{jk}\}$. The result of this very lengthy calculation is given in Appendix \ref{app-tza} where we collect all the coefficients in the above expansions for $\Omega$ and $\tilde{\Omega}$. We stress that the formulas in Appendix \ref{app-tza} are general for any modulus $k\neq0$. Using these explicit formulas, one can, for example, readily check the $YBE$ \eqref{bs:YBE_fermionic} by appropriately choosing the coefficients $\Bigl(a_{0}(u;\zeta_{j},\zeta_{k}),\ldots, a_{3}(u;\zeta_{j},\zeta_{k}),c_{1,2}(u;\zeta_{j},\zeta_{k}),d_{1,2}(u;\zeta_{j},\zeta_{k})\Bigr)$ which reduce the generic operator \eqref{tza:fermionic_R_generic_form} to the fermionic $R$-operator \eqref{bs:fermionic_R}.

We omit everywhere below the superscript $(s)$ everywhere to simplify the notations, and consider the trigonometric limit corresponding to the modulus $k=0$. We stress, however, that we let the other parameters $\zeta_{i}$ to be completely arbitrary. To further simplify our notations we also denote:
\begin{align}
    \mathcal{L}^{\alpha_{1}\alpha_{2}\alpha_{3}} \equiv \mathrm{L}^{\alpha_{1}}_{12} \mathrm{L}^{\alpha_{2}}_{13} \mathrm{L}^{\alpha_{3}}_{23}; \quad  \tilde{\mathcal{L}}^{\beta_{1}\beta_{2}\beta_{3}} \equiv \mathrm{L}^{\beta_{1}}_{23} \mathrm{L}^{\beta_{2}}_{13} \mathrm{L}^{\beta_{3}}_{12}.\label{tza:products_notation}
\end{align}
In these notations the tetrahedral relations \eqref{tza:TZA} take the form:
\begin{align}
    \mathcal{L}^{\alpha_{1}, \alpha_{2}, \alpha_{3}}=\sum_{\beta_{i}=0,1} S^{\alpha_{1} \alpha_{2} \alpha_{3}}_{\beta_{1}\beta_{2}\beta_{3}}\tilde{\mathcal{L}}^{\beta_{1}, \beta_{2}, \beta_{3}}; \quad \alpha_{1,2,3}=\{0,1\},\label{tza:TZA2}
\end{align}
One of the main results of this paper is the following solution to the tetrahedral Zamolodchikov algebraic relations \eqref{tza:TZA2}:
\begin{align*}
     \mathcal{L}^{000} &=\tilde{\mathcal{L}}^{000},\\
    \mathcal{L}^{011} &=\tilde{\mathcal{L}}^{110},\\
    \mathcal{L}^{110} &=\tilde{\mathcal{L}}^{011},\\
    \mathcal{L}^{101} &=\tilde{\mathcal{L}}^{101},\\
    \mathcal{L}^{111} &=\left(S^{111}_{010}\right)\tilde{\mathcal{L}}^{010}+\left(S^{111}_{001}\right)\tilde{\mathcal{L}}^{001}+\left(S^{111}_{100}\right)\tilde{\mathcal{L}}^{100}+\left(S^{111}_{110}\right)\tilde{\mathcal{L}}^{110}+\left(S^{111}_{011}\right)\tilde{\mathcal{L}}^{011}+\left(S^{111}_{111}\right)\tilde{\mathcal{L}}^{111},\\
    \mathcal{L}^{001} &=\left(S^{001}_{010}\right)\tilde{\mathcal{L}}^{010}+\left(S^{001}_{001}\right)\tilde{\mathcal{L}}^{001}+\left(S^{001}_{100}\right)\tilde{\mathcal{L}}^{100}+\left(S^{001}_{110}\right)\tilde{\mathcal{L}}^{110}+\left(S^{001}_{011}\right)\tilde{\mathcal{L}}^{011}+\left(S^{001}_{111}\right)\tilde{\mathcal{L}}^{111}, \\
    \mathcal{L}^{010} &=\left(S^{010}_{010}\right)\tilde{\mathcal{L}}^{010}+\left(S^{010}_{001}\right)\tilde{\mathcal{L}}^{001}+\left(S^{010}_{100}\right)\tilde{\mathcal{L}}^{100}+\left(S^{010}_{110}\right)\tilde{\mathcal{L}}^{110}+\left(S^{010}_{011}\right)\tilde{\mathcal{L}}^{011}+\left(S^{010}_{111}\right)\tilde{\mathcal{L}}^{111}, \\
    \mathcal{L}^{100} &=\left(S^{100}_{001}\right)\tilde{\mathcal{L}}^{001}+\left(S^{100}_{001}\right)\tilde{\mathcal{L}}^{001}+\left(S^{100}_{100}\right)\tilde{\mathcal{L}}^{100}+\left(S^{100}_{110}\right)\tilde{\mathcal{L}}^{110}+\left(S^{100}_{011}\right)\tilde{\mathcal{L}}^{011}+\left(S^{100}_{111}\right)\tilde{\mathcal{L}}^{111}, 
\end{align*}
where the coefficients $S^{\alpha_{1} \alpha_{2} \alpha_{3}}_{\beta_{1}\beta_{2}\beta_{3}}$ are given in Appendix \ref{app-coef}.

In addition, we note that the set of operators $\{\mathcal{L}^{\alpha_{1}\alpha_{2}\alpha_{3}}\}$ and $\{\tilde{\mathcal{L}}^{\beta_{1}\beta_{2}\beta_{3}}\}$ are linearly dependent. One can check, for example, the following relation between $\tilde{\mathcal{L}}^{\alpha_{1}\alpha_{2}\alpha_{3}}$:
\begin{align}
\tilde{\mathcal{L}}^{111}&=
\left(Y^{111}_{110}\right)\tilde{\mathcal{L}}^{110}
+\left(Y^{111}_{011}\right)\tilde{\mathcal{L}}^{011}
+\left(Y^{111}_{101}\right)\tilde{\mathcal{L}}^{101}
+\left(Y^{111}_{001}\right)\tilde{\mathcal{L}}^{001}
+\left(Y^{111}_{010}\right)\tilde{\mathcal{L}}^{010}
+\left(Y^{111}_{100}\right)\tilde{\mathcal{L}}^{100}.
\end{align}
The coefficients $Y^{111}_{\beta_{1} \beta_{2} \beta_{3}}$ are also given in Appendix \ref{app-coef}.

To summarize this section, we have found a solution which depends only on the differences $u_{ij}=u_{j}-u_{k}$ in one of the spectral parameters, while leaving the other parameters $\zeta_{i}$ completely arbitrary. This is of course the consequence of the decorated Yang-Baxter equations \eqref{tza:DYBE} having the same dependence on $u_{ij}$, and is in contrast to the solution of the tetrahedral Zamolodchikov algebra given in \cite{Shiroishi1995as,Shiroishi1995vc,Umeno1998b} where the solution for the coefficients $S^{\alpha_{1} \alpha_{2} \alpha_{3}}_{\beta_{1}\beta_{2}\beta_{3}}$ is not of the difference type. We also note that in our case the linear space spanned by $\tilde{\mathcal{L}}^{\beta_{1}, \beta_{2}, \beta_{3}}$ is 6-dimensional, as opposed to 3-dimensional case of \cite{Shiroishi1995as,Shiroishi1995vc,Umeno1998b} (see also \cite{Essler:2005bk}). As we explained in the introduction, the solution that depends on the differences of the spectral parameters is natural for taking the continuous limit and obtaining an integrable model of relativistic fermions. 

It is an interesting problem whether the solution to the tetrahedral Zamolodchikov algebraic relations \eqref{tza:TZA2} given in Appendix \ref{app-coef} can be reduced to other known solutions, e.g., given in \cite{Shiroishi1995as,Shiroishi1995vc,Umeno1998b}. One can show, that for some specific numerical values of the extra parameters $\zeta_{i}$, the general 6-dimensional linear space spanned by $\tilde{\mathcal{L}}^{\beta_{1}, \beta_{2}, \beta_{3}}$ indeed reduces to a 3-dimensional linear space. However, the coefficients $S^{\alpha_{1} \alpha_{2} \alpha_{3}}_{\beta_{1}\beta_{2}\beta_{3}}$ do not immediately match, being of the difference type for the solution given in this paper, as opposed to the solutions given in the works cited above.

\section{Extended Yang-Baxter equation}
\label{eybe}

An immediate application of our solution is the possibility, as we show below, to extend the fermionic $R$-operator and the corresponding Yang-Baxter equation, which in turn will generalize the transfer operator in \eqref{bs:tau}. The latter, as we explain below, will depend not only on $\{\zeta_{1}, \ldots, \zeta_{N}\}$, but also on some arbitrary functions of these parameters. 

To this end, we look for an extension of the fermionic $\mathrm{L}^{0,(s)}_{jk}$-operator \eqref{tza:L0} (i.e., 
 the original fermionic $R$-operator \eqref{bs:fermionic_R}) in the form:\footnote{Recall that we are considering the case $k=0$; we omitted the dependence on the modulus $k$ in the left-hand side of \eqref{eybe:extended_R}.}
\begin{align}\label{eybe:extended_R}
   \widetilde{\mathcal{R}}^{(s)}(u_{jk};\zeta_{j},\zeta_{k}):= \mathrm{L}^{0,(s)}_{jk}+c_{jk} \mathrm{L}^{1,(s)}_{jk}.
\end{align}
For the moment, the coefficients $c_{jk}$ in \eqref{eybe:extended_R} are arbitrary functions of $u_{jk},\zeta_j$ and  $\zeta_{k}$. The motivation to look for an $\widetilde{\mathcal{R}}$-operator of the form \eqref{eybe:extended_R} comes from the one-dimensional Hubbard model. Recall, that the latter corresponds to two copies of $XX$-models, together with an interaction term between the spins. In the notations \eqref{bs:notations} it has the form\cite{Essler:2005bk}:
\begin{align}
    \hat{\mathcal{H}}=-\sum_{j=1}^{N} \sum_{\,\,\,s=\uparrow \downarrow} \left(\Delta_{j+1,j (s)} +\Delta_{j,j+1 (s)}  \right) +\frac{\textrm{U}}{4} \sum_{j=1}^{N} \left(2 n_{j \uparrow} -1 \right)\left(2 n_{j \downarrow} -1 \right),\label{eybe:hubbard}
\end{align}
where $\textrm{U}$ is a constant describing the interaction strength. The key point is the structure of the $\mathrm{L}^{1,(s)}_{jk}$ \eqref{tza:L1} operator: it is readily verified (see \eqref{app:fermionic_R_id1}) that $\mathrm{L}^{1,(s)}_{jk}$ \eqref{tza:L1} produces a term $\sim \mathcal{P}^{(s)}_{jk}(2 n_{j (s)} -1)$.\footnote{Recall that $\mathcal{P}^{(s)}_{jk}:=1-n_{j(s)}-n_{k(s)}+\Delta_{jk(s)}+\Delta_{kj(s)}$ is the permutation operator for spin $s$.} Thus, provided the extended $\widetilde{\mathcal{R}}$-operator \eqref{eybe:extended_R}, one then can couple the spins as, for example, in \eqref{bs:fermionic_R_spin}, leading to the term $\sim \mathcal{P}^{(\uparrow)}_{jk}\mathcal{P}^{(\downarrow)}_{jk}(2 n_{j (\uparrow)} -1)(2 n_{ (\downarrow)} -1)$, which is precisely the term needed to obtain the interaction term in the one-dimensional Hubbard model \eqref{eybe:hubbard}.

We next set the problem of fixing the functions $c_{jk}(u_{jk},\zeta_j,\zeta_{k})$ in \eqref{eybe:extended_R}, by requiring the $YBE$  to be satisfied for the extended fermionic operator $\widetilde{\mathcal{R}}^{(s)}(u_{jk};\zeta_{j},\zeta_{k})$:
\begin{align}
     \widetilde{R}^{(s)}_{12}(u_{12};\zeta_{1},\zeta_{2})\widetilde{R}^{(s)}_{13}(u_{13};\zeta_{1},\zeta_{3})\widetilde{R}^{(s)}_{23}(u_{23};\zeta_{2},\zeta_{3})=\widetilde{R}^{(s)}_{23}(u_{23};\zeta_{2},\zeta_{3})\widetilde{R}^{(s)}_{13}(u_{13};\zeta_{1},\zeta_{3})\widetilde{R}^{(s)}_{12}(u_{12};\zeta_{1},\zeta_{2}).\label{eybe:extended_YBE_fermionic}
\end{align}
Substituting \eqref{eybe:extended_R} into \eqref{eybe:extended_YBE_fermionic} and using the solution for the tetrahedral Zamolodchikov algebra from the previous section one obtains, in principle, a set of complex equations on the $c_{jk}$-functions - a brief look at Appendix \ref{app-coef} shows the daunting task one is facing. We note, however, that the coefficients $S^{\alpha_{1} \alpha_{2} \alpha_{3}}_{\beta_{1}\beta_{2}\beta_{3}}$ are related as follows:
\newline
\begin{minipage}[t]{.3\textwidth}
%\raggedright
\begin{align*}
S^{111}_{001}&= S^{010}_{001}+Y^{111}_{001};\\
S^{111}_{011}&=S^{010}_{011}+Y^{111}_{011};\\
S^{111}_{100}&= S^{010}_{100}+Y^{111}_{100};\\
S^{111}_{101}&= S^{010}_{101}+Y^{111}_{101};\\
S^{111}_{110}&= S^{010}_{110}+Y^{111}_{110};\\
S^{111}_{010}&=S^{010}_{010}+Y^{111}_{010}-1;\\
\end{align*}
\end{minipage}% <---------------- Note the use of "%"
\begin{minipage}[t]{.3\textwidth}
%\raggedleft
\begin{align*}
S^{100}_{010}&= S^{001}_{010};\\
S^{100}_{001}&=S^{001}_{001}+1;\\
S^{100}_{011}&= S^{001}_{011};\\
S^{100}_{101}&= S^{001}_{101};\\
S^{100}_{110}&= S^{001}_{110};\\
S^{100}_{100}&= S^{001}_{100}-1;\\
\end{align*}
\end{minipage}
\begin{minipage}[t]{.3\textwidth}
\begin{align*}
S^{001}_{001}&= -S^{010}_{001};\\
S^{001}_{010}&= 1-S^{010}_{010};\\
S^{001}_{100}&= 1-S^{010}_{100};\\
S^{001}_{011}&= -S^{010}_{011};\\
S^{001}_{101}&= -S^{010}_{101};\\
S^{001}_{110}&= -S^{010}_{110}.
\end{align*}
\end{minipage}
\newline
These relations suffice to show that the resulting set of conditions on the $c_{jk}$-functions in \eqref{eybe:extended_R} are reduced to a single equation:
\begin{align}\label{eybe:cool}
    c_{23}=\frac{c_{12}-c_{13}}{c_{12} c_{13}-1}.
\end{align}
The general solution of \eqref{eybe:cool} is given by:
\begin{align}\label{eybe:sol_tanh}
    c_{jk}=\tanh\left[h_{j}(u_{j},\zeta_{j}) - h_{k}(u_{k},\zeta_{k})\right],
\end{align}
where $h_{j}(u_{j},\zeta_{j})$ are arbitrary functions.\footnote{This fact is a consequence of the modulus $k=0$ in our solution; we expect that in general, for the case $k \neq 0$, a restriction on $h_{j}(u_{j},\zeta_{j})$ should appear.} Thus, we find the following solution for the extended $ \widetilde{\mathcal{R}}^{(s)}(u_{jk};\zeta_{j},\zeta_{k})$-operator \eqref{eybe:extended_R}:
\begin{align}\label{eybe:extended_R_sol}
   \widetilde{\mathcal{R}}^{(s)}(u_{jk};\zeta_{j},\zeta_{k}):= \mathrm{L}^{0,(s)}_{jk}+\tanh\left[h_{j}(u_{j},\zeta_{j}) - h_{k}(u_{k},\zeta_{k})\right] \mathrm{L}^{1,(s)}_{jk}.
\end{align}
If the dependence on the difference $u_{jk}=u_{j}-u_{k}$ is of no interest, then one can consider the generic form of the solution in \eqref{eybe:sol_tanh} and \eqref{eybe:extended_R_sol} and investigate the consequences by considering the monodromy operator \eqref{eybe:monodromy}, the transfer matrix \eqref{eybe:transfer}, commuting quantities \eqref{eybe:tau_commute} and coupling the spins in this general setting - for each choice of $h_{j}(u_{j},\zeta_{j})$.\footnote{It will likely be the case for the one-dimensional Hubbard model.}
The monodromy and transfer operators are constructed in the usual manner (c.f. \eqref{bs:monodromy} - \eqref{bs:tau_commute}):
\begin{align}
   &\widetilde{\mathcal{T}}_{a}(u_{a},\{u_{j}\};\zeta_{a},\{\zeta_{j}\})=\widetilde{\mathcal{R}}_{aN}(u_{aN};\zeta_{a},\zeta_{N})\widetilde{\mathcal{R}}_{a,N-1}(u_{a,N-1};\zeta_{a},\zeta_{N-1})\cdot\ldots \cdot \widetilde{\mathcal{R}}_{a1}(u_{a1};\zeta_{a},\zeta_{1}),\label{eybe:monodromy}\\
       &\widetilde{{\tau}}\left(u_{a},\{u_{j}\};\zeta_{a},\{\zeta_{j}\}\right)=\str_{a}\left[ \widetilde{\mathcal{T}}_{a}\left(u_{a},\{u_{j}\};\zeta_{a},\{\zeta_{j}\}\right)\right],\label{eybe:transfer}
\end{align}
with:
\begin{align}\label{eybe:tau_commute}
    \left[\widetilde{\tau}\left(u_{a},\{u_{j}\};\zeta_{a},\{\zeta_{j}\}\right),\widetilde{\tau}\left(u_{b},\{u_{j}\};\zeta_{b},\{\zeta_{j}\}\right) \right]=0.
\end{align}
We also note here that the transfer operator $\widetilde{\tau}\left(u_{a},\{u_{j}\};\zeta_{a},\{\zeta_{j}\}\right)$ depends on two free parameters: $u_{a}$ and $\zeta_{a}$. Thus, one can explore this freedom to construct the  commuting quantities by using a more general formula:
\begin{align}
    \hat{{I}}^{(n,k)} &\sim \frac{d^{(n)}}{du_{a}^{(n)}} \frac{d^{(k)}}{d\zeta_{a}^{(k)}}\widetilde{\tau}\left(u_{a},\{u_{j}\};\zeta_{a},\{\zeta_{j}\}\right)\Big|_{\substack{u_{j} \to u_{a} \\ \zeta_{j} \to \zeta_{a}}}.\label{eybe:general_formula_v2a}
    \end{align}

On the other hand, for the extended $ \widetilde{\mathcal{R}}^{(s)}(u_{jk};\zeta_{j},\zeta_{k})$-operator \eqref{eybe:extended_R} to be of the difference type, one must, however, further restrict the $h_{j}(u_{j},\zeta_{j})$ functions and the solution \eqref{eybe:cool} as follows:
\begin{align}\label{eybe:sol_tanh_restrict}
    c_{jk}=\tanh\left[\alpha (u_{j}-u_{k})-\left\{g_{j}(\zeta_{j})-g_{k}(\zeta_{k})\right\}\right],
\end{align}
where $\alpha$ is an arbitrary constant, and $g_{j}(\zeta_{j});\,j=1,2,3$ are still arbitrary functions of $\zeta_{j}$. Thus, in this case the extended $ \widetilde{\mathcal{R}}^{(s)}(u_{jk};\zeta_{j},\zeta_{k})$-operator \eqref{eybe:extended_R} takes the form:\footnote{Note, that by using the inverse Jordan-Wigner transformation one can show that the fermionic $\widetilde{\mathcal{R}}^{(s)}(u_{jk};\zeta_{j},\zeta_{k})$-operator satisfies the free-fermion condition \eqref{bs:free_fermion_condition}.}
\begin{align}\label{eybe:extended_R_sol_restricted}
   \widetilde{\mathcal{R}}^{(s)}(u_{jk};\zeta_{j},\zeta_{k}):= \mathrm{L}^{0,(s)}_{jk}+\tanh\left[\alpha (u_{j}-u_{k})-\left\{g_{j}(\zeta_{j})-g_{k}(\zeta_{k})\right\}\right] \mathrm{L}^{1,(s)}_{jk}.
\end{align}

The above comments are also valid for the systems with coupled spins, which can be obtained by using the $ \widetilde{\mathcal{R}}^{(s)}(u_{jk};\zeta_{j},\zeta_{k})$-operator \eqref{eybe:extended_R_sol} and the corresponding $YBE$ \eqref{eybe:extended_YBE_fermionic}. The simplest case is to proceed as in \eqref{bs:fermionic_R_spin} using the general extended :
\begin{align}
   {\bm{R}}_{jk}(u_{jk};\zeta_{j},\zeta_{k};\hat{\zeta}_{j},\hat{\zeta}_{k};h_{1},h_{2},\hat{h}_{1},\hat{h}_{2}):=\widetilde{\mathcal{R}}^{(\uparrow)}_{jk}(u_{jk};\zeta_{j},\zeta_{k};h_{1},h_{2})\widetilde{\mathcal{R}}^{(\downarrow)}_{jk}(u_{jk};\hat{\zeta}_{j},\hat{\zeta}_{k};\hat{h}_{1},\hat{h}_{2}). \label{eybe:fermionic_R_spin}
\end{align}
We have written explicitly, in the above formula, the dependencies on the arbitrary functions $h_{1},h_{2}$ (for spin $\uparrow$) and $\hat{h}_{1},\hat{h}_{2}$ (for spin $\downarrow$). The ${\bm{R}}_{jk}(u_{jk};\zeta_{j},\zeta_{k};\hat{\zeta}_{j},\hat{\zeta}_{k};h_{1},h_{2},\hat{h}_{1},\hat{h}_{2})$-operator satisfies the corresponding $YBE$ obtained by taking the product of \eqref{eybe:extended_YBE_fermionic} for spin $\uparrow$ and spin $\downarrow$. The transfer operator $\widetilde{{\bm{\tau}}}(u_{a},\{u_{j}\};\zeta_{a},\{\zeta_{j}\})$ can then be defined and the local conserved quantities may be extracted as in the above case for the spinless fermions by means of the formula \eqref{eybe:general_formula_v2a}. 

To summarize, by finding the most general solution to the tetrahedral Zamolodchikov algebra in the trigonometric limit, we have found an extension for the fermionic $R$-operator of the difference type which led to an extended Yang-Baxter equation and the transfer operator depending on two free parameters. 

One can also try to construct different extensions, for example of the following type: (c.f. \cite{Korepanov:1989tl,Umeno1998,Umeno1998b,Essler:2005bk}):
\begin{align}\label{eybe:extensions2}
    \mathfrak{R}_{jk} = L^{0 (\uparrow)}_{jk} L^{0 (\downarrow)}_{jk} + \sigma_{jk} L^{1 (\uparrow)}_{jk} L^{1 (\downarrow)}_{jk}.
\end{align}
This should be now possible to analyze, using the explicit solution for the tetrahedral Zamolodchikov algebra given in Appendix \ref{app-coef}.

Another open problem is the elliptic case with $k \neq 0$. Despite being technically more complex, the task is now more feasible, given the explicit formulas in Appendix \ref{app-tza}. These and related questions will be investigated in a future publication.

\appendix
\section{Appendix}
\label{app-id}
In this appendix we list some useful identities and formulas used to calculate the tetrahedral Zamolodchikov algebra. Starting from the fermionic $R$-operator \eqref{bs:fermionic_R}:
\begin{align}
    R_{jk}(u;\zeta_{j},\zeta_{k})&=a(u;\zeta_{j},\zeta_{k})\left[-n_{j} n_{k} \right] +a(u;\zeta_{j},\zeta_{k})\left[(1-n_{j})(1- n_{k}) \right]+b(u;\zeta_{j},\zeta_{k})\left[n_{j}(1- n_{k}) \right] \nonumber\\
    &+b'(u;\zeta_{j},\zeta_{k})\left[n_{k}(1- n_{j})\right] +c(u;\zeta_{j},\zeta_{k})\left[\Delta_{jk}+\Delta_{kj} \right] +d(u;\zeta_{j},\zeta_{k})\left[-\tilde{\Delta}^{(\dagger)}_{jk}-\tilde{\Delta}_{jk}\right] ,\label{app:fermionic_R}
\end{align}
one can show the following identities used to derive the relation \eqref{tza:identity_main}, as well as the decorated Yang-Baxter equation \eqref{tza:DYBE}:
\begin{align}
    (2n_{j}-1)R_{jk}(u;\zeta_{j},\zeta_{k})  (2n_{k}-1) &=  (2n_{k}-1)R_{jk}(u;\zeta_{j},\zeta_{k})(2n_{j}-1),\\
     (2n_{j}-1)R_{jk}(u;\zeta_{j},\zeta_{k})  (2n_{j}-1) &=  (2n_{k}-1)R_{jk}(u;\zeta_{j},\zeta_{k})(2n_{k}-1).
\end{align}
In addition, one has the following relations:
\begin{align}
    R_{jk}(u;\zeta_{j},\zeta_{k})(2n_{j}-1)&=a(u;\zeta_{j},\zeta_{k})\left[-n_{j} n_{k} \right] -a(u;\zeta_{j},\zeta_{k})\left[(1-n_{j})(1- n_{k}) \right]+b(u;\zeta_{j},\zeta_{k})\left[n_{j}(1- n_{k}) \right] \nonumber\\
    &-b'(u;\zeta_{j},\zeta_{k})\left[n_{k}(1- n_{j})\right] +c(u;\zeta_{j},\zeta_{k})\left[-\Delta_{jk}+\Delta_{kj} \right] +d(u;\zeta_{j},\zeta_{k})\left[\tilde{\Delta}^{(\dagger)}_{jk}-\tilde{\Delta}_{jk}\right],\label{app:fermionic_R_id1}\\
    \nonumber \\
   (2n_{j}-1) R_{jk}(u;\zeta_{j},\zeta_{k})&=a(u;\zeta_{j},\zeta_{k})\left[-n_{j} n_{k} \right] -a(u;\zeta_{j},\zeta_{k})\left[(1-n_{j})(1- n_{k}) \right]+b(u;\zeta_{j},\zeta_{k})\left[n_{j}(1- n_{k}) \right] \nonumber\\
    &-b'(u;\zeta_{j},\zeta_{k})\left[n_{k}(1- n_{j})\right] +c(u;\zeta_{j},\zeta_{k})\left[\Delta_{jk} - \Delta_{kj} \right] +d(u;\zeta_{j},\zeta_{k})\left[-\tilde{\Delta}^{(\dagger)}_{jk}+\tilde{\Delta}_{jk}\right],
    \label{app:fermionic_R_id2}
\end{align}
\section{The equation for the tetrahedral Zamolodchikov algebra}
\label{app-tza}
Below we list the set of the coefficients in the expansions for $\Omega$ \eqref{tza:Omega} and $\tilde{\Omega}$ \eqref{tza:Omegatilde}, which are necessary to obtain the set of equations for the tetrahedral Zamolodchikov algebra. We use here the following shorthand notations:
\begin{align*}
   u&:=(u;\zeta_{1},\zeta_{3});\quad v:=(u;\zeta^{\prime}_{2},\zeta^{\prime}_{3});\quad w:=(w;\zeta^{\prime \prime}_{1},\zeta^{\prime \prime}_{2}).
\end{align*}

\subsection{\texorpdfstring{$\Omega:$}{Omega}}
\begin{align*}
\Gamma^{(0)}_{0}&=a_{0}(w)a_{0}(u)a_{0}(v)+d_{2}(w)c_{1}(u)d_{1}(v); \\
\Gamma^{(0)}_{n_{1}}&= a_{0}(v)\left\{a_{0}(w)a_{1}(u)+a_{1}(w)a_{0}(u) + a_{1}(w)a_{1}(u)\right\}+c_{1}(w)d_{2}(u)d_{1}(v)-d_{2}(w)c_{1}(u)d_{1}(v);\\
\Gamma^{(0)}_{n_{2}}&= a_{0}(u)\left\{a_{0}(w)a_{1}(v)+a_{2}(w)a_{0}(v) + a_{2}(w)a_{1}(v)\right\}+c_{2}(w)c_{1}(u)c_{2}(v)-d_{2}(w)c_{1}(u)d_{1}(v);\\
\Gamma^{(0)}_{n_{3}}&= a_{0}(w)\left\{a_{0}(u)a_{2}(v)+a_{2}(u)a_{0}(v) + a_{2}(u)a_{2}(v)\right\}-d_{2}(w)c_{1}(u)d_{1}(v)+d_{2}(w)d_{1}(u)c_{1}(v);\\
\Gamma^{(0)}_{n_{1} n_{2}}&= a_{0}(w)a_{1}(u)a_{1}(v)+a_{1}(w)a_{1}(v)\left\{a_{0}(u)+a_{1}(u)\right\}-c_{1}(w)d_{2}(u)d_{1}(v)-c_{2}(w)c_{1}(u)c_{2}(v)\\
 &\left\{ a_{2}(w)a_{1}(u) + a_{3}(w)a_{0}(u) +a_{3}(w)a_{1}(u) \right\} \left\{a_{0}(v)+a_{1}(v)\right\}
 - d_{1}(w)d_{2}(u)c_{2}(v) + d_{2}(w)c_{1}(u)d_{1}(v);\\
 \Gamma^{(0)}_{n_{1} n_{3}}&= \left\{a_{0}(w) +a_{1}(w)\right\} \left\{a_{1}(u)a_{2}(v)+a_{3}(u)a_{0}(v)+a_{3}(u)a_{2}(v)\right\}-c_{1}(w)c_{2}(u)c_{1}(v)-c_{1}(w)d_{2}(w)d_{1}(v)\\
 &+a_{1}(w)\left\{a_{0}(u)a_{2}(v)+a_{2}(u)a_{0}(v)+a_{2}(u)a_{2}(v)\right\}- d_{2}(w)c_{1}(u)d_{1}(v) - d_{2}(w)d_{1}(u)c_{1}(v);\\
 \Gamma^{(0)}_{n_{2} n_{3}}&= \left\{a_{0}(w)+a_{2}(w)\right\}\left\{a_{0}(u)a_{3}(v)+a_{2}(u)a_{1}(v)+a_{2}(u)a_{3}(v)\right\}-c_{2}(w)c_{1}(u)c_{2}(v)-c_{2}(w)d_{1}(u)d_{2}(v)\\
 &+a_{2}(w)\left\{a_{0}(u)a_{2}(v)+a_{2}(u)a_{0}(v)+a_{2}(u)a_{2}(v)\right\}+ d_{2}(w)c_{1}(u)d_{1}(v) - d_{2}(w)d_{1}(u)c_{1}(v);\\
\Gamma^{(0)}_{n_{1} n_{2} n_{3}}&= a_{0}(w)\left\{a_{1}(u)a_{3}(v)+a_{3}(u)a_{1}(v)+a_{3}(u)a_{3}(v)\right\}+a_{1}(w)\left\{a_{0}(u)a_{3}(v)+a_{2}(u)a_{1}(v)+a_{2}(u)a_{3}(v)\right.\\
 &\left. +a_{1}(u)a_{3}(v) + a_{3}(u)a_{1}(v) +a_{3}(u)a_{3}(v)\right\}+ a_{2}(w)\left\{a_{1}(u)a_{2}(v)+a_{3}(u)a_{0}(v)+a_{3}(u)a_{2}(v)\right.\\
 &\left.+a_{1}(u)a_{3}(v) + a_{3}(u)a_{1}(v) +a_{3}(u)a_{3}(v)\right\}+a_{3}(w)\left\{a_{0}(u)a_{2}(v)+v)a_{2}(u)a_{0}(v)+a_{2}(u)a_{2}(v)\right.\\
 &\left.+a_{1}(u)a_{2}(v) +a_{3}(u)a_{0}(v)+a_{3}(u)a_{2}(v)+a_{0}(u)a_{3}(v) +a_{2}(u)a_{1}(v) +a_{2}(u)a_{3}(v) +a_{1}(u)a_{3}(v)\right.\\
 &\left.+a_{3}(u)a_{1}(v)+a_{3}(u)a_{3}(v) \right\} +c_{1}(w)c_{2}(u)c_{1}(v) +c_{1}(w)d_{2}(u)d_{1}(v)+c_{2}(w)c_{1}(u)c_{2}(v)\\
 &+c_{2}(w)d_{1}(u)d_{2}(v) +d_{1}(w)d_{2}(u)c_{2}(v)-d_{1}(w)c_{2}(u)d_{2}(v)+d_{2}(w)d_{1}(u)c_{1}(v)\\
 &-d_{2}(w)c_{1}(u)c_{1}(v);\\
 \Gamma^{(1)}_{\Delta_{23}}&=a_{0}(u)c_{1}(v)\left\{a_{0}(w)+ a_{2}(w)\right\} +c_{2}(w)c_{1}(u)\left\{a_{0}(v) + a_{2}(v)\right\};\\
  \Gamma^{(1)}_{\tilde{\Delta}^{(+)}_{23}}&=
  d_{1}(v)\left\{a_{0}(u) + a_{2}(u)\right\} \left\{a_{0}(w)+ a_{2}(w)\right\} +c_{2}(w)d_{1}(u)a_{0}(v);\\
  \Gamma^{(1)}_{\tilde{\Delta}_{23}}&=a_{0}(w)a_{0}(u)d_{2}(v) - d_{2}(w)c_{1}(u)\left\{a_{0}(v)+a_{1}(v)+a_{2}(v)+a_{3}(v)\right\};\\
  \Gamma^{(1)}_{{\Delta}_{13}}&=c_{1}(u)\left\{a_{0}(w) +a_{1}(w) \right\} \left\{ a_{0}(v) + a_{2}(v)\right\} +c_{1}(w)a_{0}(u)c_{1}(v);\\
  \Gamma^{(1)}_{{\Delta}_{31}}&=a_{0}(w)c_{2}(u)a_{0}(v) -d_{2}(w)d_{1}(v)\left\{ a_{0}(u)+a_{1}(u)+a_{2}(u)+a_{3}(u)\right\};\\
   \Gamma^{(1)}_{\tilde{\Delta}^{(+)}_{13}}&=a_{0}(v)d_{1}(u) \left\{ a_{0}(w)+ a_{1}(w)\right\} +c_{1}(w)d_{1}(v)\left\{a_{0}(u) + a_{2}(u \right\};\\
   \Gamma^{(1)}_{\tilde{\Delta}_{13}}&=a_{0}(w)d_{2}(u)\left\{ a_{0}(v)+a_{2}(v) \right\} +d_{2}(w)c_{1}(v)\left\{ a_{0}(u) + a_{1}(u)\right\};\\
    \Gamma^{(1)}_{n_{1} {\Delta}_{23}}&= a_{1}(u)c_{1}(v)\left\{a_{0}(w)+a_{1}(w)+a_{2}(w)+a_{3}(w) \right\} +a_{0}(u)c_{1}(v)\left\{a_{1}(w)+a_{3}(w)\right\}\\
    &-\left\{c_{2}(w)c_{1}(u)+d_{1}(w)d_{2}(u)\right\}\left\{a_{0}(v)+a_{2}(v)\right\};\\
    \Gamma^{(1)}_{n_{1} {\tilde{\Delta}}_{23}}&=
    \left\{a_{0}(w) +a_{1}(w) \right\}a_{1}(u)d_{2}(v)+a_{1}(w)a_{0}(u)d_{2}(v)+\left\{d_{2}(w)c_{1}(u)
    -c_{1}(w)d_{2}(u)\right\} \times\\
    &\times \left\{a_{0}(v)+a_{1}(v)+a_{2}(v)+a_{3}(v) \right\};\\
            \displaybreak \\ 
    \Gamma^{(1)}_{n_{1} {\tilde{\Delta}}^{(+)}_{23}}&=d_{1}(v)\left\{a_{1}(u) +a_{3}(u) \right\} \left\{a_{0}(w)+a_{1}(w)+a_{2}(w)+a_{3}(w)\right\} + d_{1}(w)c_{2}(u)a_{0}(v)\\
    &+d_{1}(v)\left\{a_{0}(u)+a_{2}(u)\right\} \left\{a_{1}(w)+a_{3}(w)\right\} -c_{2}(w)d_{1}(u)a_{0}(v);\\
     \Gamma^{(1)}_{n_{1} {{\Delta}}_{32}}&=c_{2}(v)\left\{a_{1}(u)+a_{3}(u)\right\}\left\{a_{0}(w)+a_{1}(w)\right\} +a_{1}(w)c_{2}(v)\left\{a_{0}(u)+a_{2}(u)\right\}\\
     &-\left\{c_{1}(w)c_{2}(u)+d_{2}(w)d_{1}(u) \right\}\left\{a_{0}(v)+a_{1}(v)\right\};\\
    \Gamma^{(1)}_{n_{2} {{\Delta}}_{13}}&=c_{1}(u)\left\{a_{0}(w)+a_{1}(w)+a_{2}(w)+a_{3}(w)\right\} +c_{1}(u)\left\{a_{0}(v) +a_{2}(v)\right\} \left\{a_{2}(w)+a_{3}(w)\right\}\\
    &-c_{1}(w)a_{0}(u)c_{1}(v)+d_{1}(w)a_{0}(u)d_{2}(v);\\
    \Gamma^{(1)}_{n_{2} {{\Delta}}_{31}}&=c_{2}(u)a_{1}(v)\left\{a_{0}(w)+a_{2}(w)\right\}+
    a_{2}(w)c_{2}(u)a_{0}(v)+\left\{d_{2}(w)d_{1}(v)-c_{2}(w)c_{2}(v)\right\}\times \\
    &\times\left\{a_{0}(u)+a_{1}(u)+a_{2}(u)+a_{3}(u)\right\};\\
    \Gamma^{(1)}_{ {n_{2} \tilde{\Delta}}^{(+)}_{13}}&=d_{1}(u)a_{1}(v)\left\{a_{0}(w)+a_{1}(w)+a_{2}(w)+a_{3}(w) \right\}+d_{1}(u)a_{0}(v)\left\{a_{2}(w)+a_{3}(w)\right\}\\
    &-c_{1}(w)d_{1}(v)\left\{a_{0}(u)+a_{2}(u)\right\}-d_{1}(w)c_{2}(v)\left\{a_{0}(u)+a_{2}(u)\right\};\\
     \Gamma^{(1)}_{ {n_{2} \tilde{\Delta}}_{13}}&=d_{2}(u)\left\{a_{1}(v)+a_{3}(v)\right\}\left\{a_{0}(w)+a_{2}(w)\right\} +a_{2}(w)d_{2}(u)\left\{a_{0}(v)+a_{2}(v)\right\}\\ 
     &-\left\{c_{2}(w)d_{2}(v)+d_{2}(w)c_{1}(v)\right\}\left\{a_{0}(u)+a_{1}(u)\right\};\\
     \Gamma^{(1)}_{{\Delta}_{12}}&=c_{1}(u)c_{2}(v)\left\{a_{0}(w)+a_{1}(w)\right\} +c_{1}(w)a_{0}(u)\left\{a_{0}(v)+a_{1}(v)\right\};\\
     \Gamma^{(1)}_{{n_{3} {\Delta}}_{12}}&= -\left\{c_{1}(u)c_{2}(v)+d_{1}(u)d_{2}(v)\right\} \left\{a_{0}(w)+a_{1}(w)\right\}+c_{1}(w)a_{0}(u)\left\{a_{2}(v)+a_{3}(v)\right\}\\
     &+c_{1}(w)a_{2}(u)\left\{a_{0}(v)+a_{1}(v)+a_{2}(v)+a_{3}(v) \right\};\\
     \Gamma^{(1)}_{{\tilde{\Delta}}^{(+)}_{12}}&= -c_{1}(u)d_{1}(v)\left\{a_{0}(w)+a_{1}(w)+a_{2}(w)+a_{3}(w) \right\} +a_{0}(u)a_{0}(v)d_{1}(w);\\
     \Gamma^{(1)}_{n_{3}{\tilde{\Delta}}^{(+)}_{12}}&=\left\{c_{1}(u)d_{1}(v) - d_{1}(u)c_{1}(v)\right\} \left\{a_{0}(w)+a_{1}(w)+a_{2}(w)+a_{3}(w) \right\} +d_{1}(w)a_{0}(u)a_{2}(v)\\
     &+d_{1}(w)a_{2}(u)\left\{a_{0}(v)+a_{2}(v)\right\};\\
    \Gamma^{(1)}_{{\tilde{\Delta}}_{12}}&=a_{0}(w)d_{2}(u)c_{2}(v)+\left\{d_{2}(w)a_{0}(v)+d_{2}(w)a_{1}(v)\right\}\left\{a_{0}(u)+a_{1}(v)\right\};\\
     \Gamma^{(1)}_{n_{3}{\tilde{\Delta}}_{12}}&=a_{0}(w)
\left\{c_{2}(u)d_{2}(v) - d_{2}(u)c_{2}(v)\right\}+d_{2}(w)\left\{a_{0}(v)+a_{1}(v)\right\}\left\{a_{2}(u)+a_{3}(u)\right\}\\
&+d_{2}(w)\left\{a_{2}(v)+a_{3}(v)\right\}\left\{a_{0}(u)+a_{1}(u)+a_{2}(u)+a_{3}(u) \right\};\\
 \Gamma^{(1)}_{{{\Delta}}_{21}}&=d_{2}(u)d_{1}(v)\left\{a_{0}(w)+a_{2}(w)\right\}+c_{2}(w)a_{0}(v)\left\{a_{0}(u)+a_{1}(u)\right\};\\
  \Gamma^{(1)}_{{n_{3} {\Delta}}_{21}}&= -\left\{d_{2}(u)d_{1}(v)+c_{2}(u)c_{1}(v)\right\} \left\{a_{0}(w)+a_{2}(w)\right\} +c_{2}(w)a_{2}(v)\left\{a_{0}(u)+a_{1}(u)\right\}\\
  &+c_{2}(w)\left\{a_{0}(v)+a_{2}(v)\right\}\left\{a_{2}(u)+a_{3}(v)\right\};\\
   \Gamma^{(1)}_{{{\Delta}}_{32}}&=a_{0}(w)c_{2}(v)\left\{a_{0}(u)+a_{2}(u)\right\} + d_{2}(w)d_{1}(u)\left\{a_{0}(v)+a_{1}(v)\right\}.
\end{align*}
\newpage
\subsection{\texorpdfstring{$\tilde{\Omega}:$}{Omega tilde}}
\label{subsection:Omegatilde}
\begin{align*}
\tilde{\Gamma}^{(0)}_{0}&=a_{0}(v)a_{0}(u)a_{0}(w)+d_{2}(v)c_{2}(u)d_{1}(w); \\
\tilde{\Gamma}^{(0)}_{n_{1}}&= a_{0}(v)\left\{a_{1}(u)a_{0}(w)+a_{0}(u)a_{1}(w) + a_{1}(u)a_{1}(w)\right\}+d_{2}(v)d_{1}(u)c_{2}(w)-d_{2}(v)c_{2}(u)d_{1}(w);\\
\tilde{\Gamma}^{(0)}_{n_{2}}&= a_{0}(u)\left\{a_{1}(v)a_{0}(w)+a_{0}(v)a_{2}(w) + a_{1}(v)a_{2}(w)\right\}+c_{1}(v)c_{2}(u)c_{1}(w)-d_{2}(v)c_{2}(u)d_{1}(w);\\
\tilde{\Gamma}^{(0)}_{n_{3}}&= a_{0}(w)\left\{a_{0}(v)a_{2}(u)+a_{2}(v)a_{0}(u) + a_{2}(v)a_{2}(u)\right\}-d_{2}(v)c_{2}(u)d_{1}(w)+c_{2}(w)d_{2}(u)d_{1}(w);\\
\tilde{\Gamma}^{(0)}_{n_{1} n_{2}}&= a_{1}(v)a_{1}(u)a_{0}(w)+a_{1}(v)a_{1}(w)\left\{a_{0}(u)+a_{1}(u)\right\}-c_{1}(v)c_{2}(u)c_{1}(w) -c_{1}(v)d_{1}(u)d_{2}(w)\\
&\left\{a_{0}(v)+a_{1}(v)\right\} \left\{a_{1}(u)a_{2}(w)+a_{0}(u)a_{3}(w)+a_{1}(u)a_{3}(w)\right\}+d_{2}(v)c_{2}(u)d_{1}(w) - d_{2}(v)d_{1}(u)c_{2}(w);\\
 \tilde{\Gamma}^{(0)}_{n_{1} n_{3}}&= \left\{a_{0}(w) +a_{1}(w)\right\} \left\{a_{0}(v)a_{3}(u)+a_{2}(v)a_{1}(u)+a_{2}(v)a_{3}(u)\right\}-c_{2}(v)c_{1}(u)c_{2}(w) -c_{2}(v)d_{2}(u)d_{1}(w)\\
&+a_{1}(w)\left\{a_{0}(v)a_{2}(u)+a_{2}(v)a_{0}(u)+a_{2}(v)a_{2}(u)\right\}+ d_{2}(v)c_{2}(u)d_{1}(w)-d_{2}(v)d_{1}(u)c_{2}(w);\\
 \tilde{\Gamma}^{(0)}_{n_{2} n_{3}}&= \left\{a_{0}(w)+a_{2}(w)\right\} \left\{a_{1}(v)a_{2}(u)+a_{3}(v)a_{0}(u)+a_{3}(u)a_{2}(v)\right\}-c_{1}(v)c_{2}(u)c_{1}(w) -c_{2}(v)d_{2}(u)d_{1}(w)\\
 &+a_{2}(w)\left\{a_{0}(v)a_{2}(u)+a_{2}(v)a_{0}(u)+a_{2}(v)a_{2}(u)\right\}-d_{1}(v)d_{2}(u)c_{1}(w) + d_{2}(v)c_{2}(u)d_{1}(w);\\
\tilde{\Gamma}^{(0)}_{n_{1} n_{2} n_{3}}&= a_{0}(w)\left\{a_{1}(v)a_{3}(u)+a_{3}(v)a_{1}(u)+a_{3}(v)a_{3}(u)\right\}+a_{1}(w)\left\{a_{1}(v)a_{2}(u)+a_{3}(v)a_{0}(u)+a_{3}(v)a_{2}(u)\right.\\
 &\left. +a_{1}(v)a_{3}(u) + a_{3}(v)a_{1}(u) +a_{3}(v)a_{3}(u)\right\}+ a_{2}(w)\left\{a_{0}(v)a_{3}(u)+a_{2}(v)a_{1}(u)+a_{2}(v)a_{3}(u)\right.\\
 &\left.+a_{1}(v)a_{3}(u) + a_{3}(v)a_{1}(u) +a_{3}(v)a_{3}(u)\right\}+a_{3}(w)\left\{a_{0}(v)a_{2}(u)+a_{2}(v)a_{0}(u)+a_{2}(v)a_{2}(u)\right.\\
 &\left.+a_{1}(v)a_{2}(u) +a_{3}(v)a_{0}(u)+a_{3}(v)a_{2}(u)+a_{0}(v)a_{3}(u) +a_{2}(v)a_{1}(u) +a_{2}(v)a_{3}(u) +a_{1}(v)a_{3}(u)\right.\\
 &\left.+a_{3}(v)a_{1}(u)+a_{3}(v)a_{3}(u) \right\} +c_{1}(v)c_{2}(u)c_{1}(w) +d_{1}(v)d_{2}(u)c_{1}(w)+d_{2}(v)d_{1}(u)c_{2}(w)\\
 &+c_{2}(v)c_{1}(u)c_{2}(w) +c_{2}(v)d_{2}(u)d_{1}(w)-d_{2}(v)c_{2}(u)d_{1}(w)+c_{1}(v)d_{1}(u)d_{2}(w)\\
 &-d_{1}(v)c_{1}(u)d_{2}(w);\\
 \tilde{\Gamma}^{(1)}_{\Delta_{23}}&=a_{0}(w)c_{1}(v)\left\{a_{0}(u)+ a_{2}(u)\right\} +d_{2}(u)d_{1}(w)\left\{a_{0}(v) + a_{1}(v)\right\};\\
  \tilde{\Gamma}^{(1)}_{\tilde{\Delta}^{(+)}_{23}}&=
  d_{1}(v)a_{0}(u)a_{0}(w) - d_{1}(w)c_{2}(u)\left\{a_{0}(v)+a_{1}(v)+a_{2}(v)+a_{3}(v)\right\};\\
  \tilde{\Gamma}^{(1)}_{\tilde{\Delta}_{23}}&= d_{2}(v)\left\{a_{0}(u)+a_{2}(u)\right\} \left\{a_{0}(w)+a_{2}(w) \right\} + a_{0}(v)d_{2}(u)c_{1}(w);\\
  \tilde{\Gamma}^{(1)}_{{\Delta}_{13}}&=a_{0}(v)c_{1}(u)a_{0}(w) - d_{2}(v)d_{1}(w)\left\{a_{0}(u)+a_{1}(u)+a_{2}(u)+a_{3}(u)\right\};\\
  \tilde{\Gamma}^{(1)}_{{\Delta}_{31}}&=c_{2}(u)\left\{a_{0}(v)+c_{2}(v)\right\} \left\{a_{0}(w) + a_{1}(w) \right\} + c_{2}(v)a_{0}(u)c_{2}(w);\\
\tilde{\Gamma}^{(1)}_{\tilde{\Delta}^{(+)}_{13}}&=a_{0}(w)d_{1}(u)\left\{a_{0}(v) + a_{2}(v)\right\} +c_{2}(v)d_{1}(w)\left\{a_{0}(u) +a_{1}(u) \right\};\\
   \tilde{\Gamma}^{(1)}_{\tilde{\Delta}_{13}}&=a_{0}(v)d_{2}(u)\left\{a_{0}(w)+a_{1}(w) \right\} +d_{2}(v)c_{2}(w)\left\{a_{0}(u) +a_{2}(u)\right\};\\
   \tilde{\Gamma}^{(1)}_{n_{1} {\Delta}_{23}}&= c_{1}(v)\left\{a_{1}(u)+a_{3}(u)\right\} \left\{a_{0}(w)+a_{1}(w) \right\} -c_{1}(u)c_{2}(w)\left\{a_{0}(v)+a_{1}(v)\right\}\\
    &-d_{2}(u)d_{1}(w)\left\{a_{0}(v)+a_{1}(v) \right\}+c_{1}(v)a_{1}(w)\left\{a_{0}(u)+a_{2}(u)\right\};
    \displaybreak \\
    \tilde{\Gamma}^{(1)}_{n_{1} {\tilde{\Delta}}_{23}}&= d_{2}(v)\left\{a_{1}(u)+a_{3}(u) \right\} \left\{a_{0}(w)+a_{1}(w)+a_{2}(w)+a_{3}(w) \right\}+ a_{0}(v)c_{1}(u)d_{2}(w)\\
    &+d_{2}(v)\left\{a_{0}(u)+a_{2}(u)\right\} \left\{a_{1}(w)+a_{3}(w) \right\} - a_{0}(v)d_{2}(u)c_{1}(w);\\
    \tilde{\Gamma}^{(1)}_{n_{1} {\tilde{\Delta}}^{(+)}_{23}}&=d_{1}(v)a_{1}(u)\left\{a_{0}(w)+a_{1}(w)\right\} +d_{1}(v)a_{0}(u)a_{1}(w)\\
    &+\left\{d_{1}(w)c_{2}(u)-c_{2}(w)d_{1}(u)\right\}\left\{a_{0}(v)+a_{1}(v)+a_{2}(v)+a_{3}(v)\right\};\\
    \tilde{\Gamma}^{(1)}_{n_{1} {{\Delta}}_{32}}&=c_{2}(v)a_{1}(u)\left\{a_{0}(w)+a_{1}(w)+a_{2}(w)+a_{3}(w)\right\} +c_{2}(v)a_{0}(u)\left\{a_{1}(w) +a_{3}(w) \right\}\\
    &-\left\{c_{2}(u)c_{1}(w)+d_{1}(u)d_{2}(w)\right\}\left\{a_{0}(v)+a_{2}(v) \right\};\\
    \tilde{\Gamma}^{(1)}_{{n_{2} {\Delta}}_{13}}&= a_{1}(v)c_{1}(u)\left\{ a_{0}(w) +a_{2}(w)\right\} +a_{2}(w)a_{0}(v)c_{1}(u)\\
    &+\left\{d_{1}(w)d_{2}(v) -c_{1}(v)c_{1}(w) \right\} \left\{ a_{0}(u)+a_{1}(u)+a_{2}(u)+a_{3}(u)\right\};\\
     \tilde{\Gamma}^{(1)}_{ {n_{2} {\Delta}}_{31}}&=c_{2}(u)\left\{a_{1}(v)+a_{3}(v) \right\}\left\{a_{0}(w)+a_{1}(w)+a_{2}(w)+a_{3}(w) \right\}-c_{2}(v)a_{0}(u)c_{2}(w)\\
     &+c_{2}(u)\left\{a_{0}(v)+a_{2}(v) \right\}\left\{a_{2}(w)+a_{3}(w) \right\}+d_{1}(v)a_{0}(u)d_{2}(w);\\
     \tilde{\Gamma}^{(1)}_{n_{2}{\tilde{\Delta}}^{(+)}_{13}}&= d_{1}(u)\left\{a_{1}(v)+a_{3}(v) \right\} \left\{a_{0}(w)+a_{2}(w) \right\} +d_{1}(u)a_{2}(w)\left\{a_{0}(v)+a_{2}(v)\right\}\\
     &-\left\{c_{1}(w)d_{1}(v) + d_{1}(w)c_{2}(v)\right\}\left\{a_{0}(u)+a_{1}(u) \right\};\\
    \tilde{\Gamma}^{(1)}_{n_{2}{\tilde{\Delta}}_{13}}&=a_{1}(v)d_{2}(u)\left\{a_{0}(w)+a_{1}(w)+a_{2}(w)+a_{3}(w) \right\} +a_{0}(v)d_{2}(u)\left\{a_{2}(w) +a_{3}(w) \right\}\\
    &-\left\{ d_{2}(v)c_{2}(w)+c_{1}(v)d_{2}(w)\right\}\left\{a_{0}(u)+a_{2}(u)\right\};\\
 \tilde{\Gamma}^{(1)}_{{{\Delta}}_{12}}&=d_{2}(v)d_{1}(u)\left\{ a_{0}(w)+a_{2}(w)\right\} +c_{1}(w)a_{0}(v)\left\{ a_{0}(u)+a_{1}(u)\right\};\\
  \tilde{\Gamma}^{(1)}_{{n_{3} {\Delta}}_{12}}&= -\left\{c_{2}(v)c_{1}(u) +d_{2}(v)d_{1}(u) \right\}\left\{a_{0}(w)+a_{2}(w) \right\} +c_{1}(w)a_{0}(v)\left\{a_{2}(u)+a_{3}(u) \right\}\\
  &+c_{1}(w)a_{2}(v)\left\{a_{0}(u)+a_{1}(u)+a_{2}(u)+a_{3}(u) \right\};\\
     \tilde{\Gamma}^{(1)}_{{\tilde{\Delta}^{(+)}}_{12}}&=c_{1}(v)d_{1}(u)a_{0}(w)+d_{1}(w)\left\{a_{0}(u)+a_{1}(u)\right\}\left\{a_{0}(v)+a_{1}(v) \right\};\\
   \tilde{\Gamma}^{(1)}_{n_{3}{\tilde{\Delta}^{(+)}}_{12}}&= a_{0}(w)\left\{d_{1}(v)c_{1}(u)-c_{1}(v)d_{1}(u)\right\}+d_{1}(w)\left\{a_{0}(u) +a_{1}(u) \right\} \left\{a_{2}(v)+a_{3}(v) \right\}\\
   &+d_{1}(w)a_{2}(u)\left\{a_{0}(v) +a_{1}(v)+a_{2}(v)+a_{3}(v) \right\};\\
     \tilde{\Gamma}^{(1)}_{{\tilde{\Delta}}_{12}}&= a_{0}(v)a_{0}(u)d_{2}(w) -d_{2}(v)c_{2}(u)\left\{a_{0}(w)+a_{1}(w) +a_{2}(w) +a_{3}(w) \right\};\\
\tilde{\Gamma}^{(1)}_{n_{3}{\tilde{\Delta}}_{12}}&= \left\{d_{2}(v)c_{2}(u) - c_{2}(v)d_{2}(u) \right\} \left\{a_{0}(w)+a_{1}(w)+a_{2}(w) +a_{3}(w)\right\}+d_{2}(w)a_{0}(v)a_{2}(u)\\
&+d_{2}(w)a_{2}(v)\left\{a_{0}(u) +a_{2}(u) \right\};\\
\tilde{\Gamma}^{(1)}_{{{\Delta}}_{21}}&=c_{1}(v)c_{2}(u)\left\{a_{0}(w)+a_{1}(w) \right\} +c_{2}(w)a_{0}(u)\left\{a_{0}(v)+a_{1}(v) \right\};\\
\tilde{\Gamma}^{(1)}_{n_{3} {{\Delta}}_{21}}&= -\left\{c_{1}(v)c_{2}(u) +d_{1}(v)d_{2}(u) \right\} \left\{a_{0}(w) +a_{1}(w) \right\} +c_{2}(w)a_{0}(u)\left\{a_{2}(v)+a_{3}(v) \right\} \\
&+c_{2}(w)a_{2}(u)\left\{ a_{0}(v)+a_{1}(v)+a_{2}(v)+a_{3}(v)\right\};\\
\tilde{\Gamma}^{(1)}_{n_{3} {{\Delta}}_{21}}&=c_{2}(v)a_{0}(u)\left\{a_{0}(w)+a_{2}(w) \right\}+c_{2}(u)c_{1}(w)\left\{a_{0}(v)+a_{2}(v) \right\}.
\end{align*}

\newpage
\section{The list of coefficients for the Zamolodchikov algebra}
\label{app-coef}
In this appendix we list the set of all non-zero coefficients of the tetrahedral Zamolodchikov algebra $S^{abc}_{ijk}$, as well as the coefficients for the linear dependency $Y^{111}_{ijk}$. To make our formulas more readable, we use the following shorthand notations: $\co(x) \equiv \cos(x);\, \s(x)=\sin(x)$:
\begin{align*}
S^{010}_{001}&=-\frac{1}{4\mathcal{N}}\Bigl[\s(2 (u_{23}-{\zeta_{1}}))-\s(2 ({u_{23}}+{\zeta_{1}}))+8 \co({\zeta_{3}}) \s(u_{12}) \left\{\s({\zeta_{2}}) \s u_{13}-\s ({\zeta_{1}}) \s(u_{23})\right\}-2 \s(2 {\zeta_{2}}) \co(2u_{13})\\
&\left.+4 \s({\zeta_{1}}+{\zeta_{2}}) \Bigl(\co(u_{13}+u_{23})+\co(2 {\zeta_{3}}) \left\{\co({\zeta_{1}}-{\zeta_{2}})-\co(u_{12})\right\}-2 \s({\zeta_{2}}) \s({\zeta_{3}}) \co(u_{13})+2 \s({\zeta_{1}}) \s({\zeta_{3}}) \co(u_{23})\right.\\
& +\s(2 {\zeta_{3}}) \s({\zeta_{1}}-{\zeta_{2}})\Bigr)\Bigr],\\
S^{010}_{010}&=\frac{1}{2\mathcal{N}}\Bigl[-4 \s(\zeta_{2}) \left\{\co(u_{13})+\co(\zeta_{1}-\zeta_{3})\right\} \left\{\co(\zeta_{1}) \co(u_{23})-\co(\zeta_{3}) \co(u_{12})\right\}+\s(2 \zeta_{3}) \left\{\co(2 \zeta_{1})-\co(2 u_{12})\right\}\nonumber\\
&-4 \s^2(\zeta_{2}) \co(u_{13}) \s(\zeta_{1}-\zeta_{3})+\s(2 \zeta_{1}) \left\{\co(2 u_{23})\right\}-\co(2 \zeta_{3}))\Bigr],\\
S^{010}_{011}&=-\frac{\mi}{2\mathcal{N}}\Bigl[2\s(\zeta_{2})\left\{ 2\co(\zeta_{1})\left(\co(u_{23})+\co(\zeta_{2}+\zeta_{3})\right)\s(u_{13})-\co(\zeta_{3})\left(\s(u_{12}+u_{13})+\co(2\zeta_{1})\s(u_{23})\right)\right\}\\
&\left.-2\s(\zeta_{3})\left\{ 2\co(\zeta_{1})\left(\co(u_{23})+\co(\zeta_{2}+\zeta_{3})\right)\s(u_{12})+\co(\zeta_{2})\left(-\s(u_{12}+u_{13})+\co(2\zeta_{1})\s(u_{23})\right)\right\}
\right.\\
&-\s(2u_{13})\s(2\zeta_{2})+\s(2u_{12})\s(2\zeta_{3})\Bigr],\\
S^{010}_{100}&=\frac{1}{4\mathcal{N}}\Bigl[
-2\co(2\zeta_{3})\s(2\zeta_{1})+\sin(2(u_{13}-\zeta_{2}))-2\s(u_{12}-\zeta_{1}-\zeta_{2})+\s(u_{13}+u_{23}-\zeta_{1}-\zeta_{2})\\
&+\s(u_{13}+u_{23}+\zeta_{1}-\zeta_{2})-\s(2(u_{13}+\zeta_{2}))-\s(u_{13}+u_{23}-\zeta_{1}+\zeta_{2})+2\s(2(\zeta_{1}+\zeta_{2}))\\
&+2\s(u_{12}+\zeta_{1}+\zeta_{2})-\s(u_{13}+u_{23}+\zeta_{1}+\zeta_{2})+\s(2(u_{12}-\zeta_{3}))-2\s(u_{12}+u_{13}-\zeta_{2}-\zeta_{3})\\
&-8\co(\zeta_{1})\s(u_{12})\s(u_{23})\s(\zeta_{3})-\s(2(u_{12}+\zeta_{3}))-4\s(\zeta_{1})\left\{\co(u_{12})\co(\zeta_{2}+2\zeta_{3})+2\co(u_{13})\s(\zeta_{2})\s(\zeta_{2}+\zeta_{3})\right\}\\
&+2\s(u_{12}+u_{13}+\zeta_{2}+\zeta_{3})+2\co(2\zeta_{1})
\left\{\s(u_{23}-\zeta_{2}-\zeta_{3})+\s(2\zeta_{3})-\s(u_{23}+\zeta_{2}+\zeta_{3})\right\}
\Bigr],\\
S^{010}_{101}&=\frac{\mi}{2\mathcal{N}}\Bigl[ 4 \co(\zeta_{2}) \left\{\co(u_{13})+\co(\zeta_{1}-\zeta_{3})\right\} \left\{\s(\zeta_{3}) \s(u_{12})-\s (\zeta_{1}) \s(u_{23})\right\}-2 \s(\zeta_{1}+\zeta_{3}) \s (u_{12}-u_{23})\\
&-\s(2 \zeta_{3}) \s(2 (u_{12}))+\s(2 \zeta_{1}) \s(2 (u_{23}))+2 \co(2 \zeta_{2}) \s(u_{13}) \s(\zeta_{1}-\zeta_{3})\Bigr],\\
S^{010}_{110}&=\frac{\mi}{\mathcal{N}}\Bigl[8 \co (\zeta_{2}) \left\{\co(u_{12})+\co(\zeta_{1}+\zeta_{2})\right\} \left\{\s(\zeta_{2}) \s(u_{13})-\s(\zeta_{1}) \s(u_{23})\right\}+\co(2 (u_{23}-\zeta_{1}))-\co(2 (u_{23}+\zeta_{1}))\\
&-\co(2 (u_{13}-\zeta_{2}))+\co(2 (u_{13}+\zeta_{2}))-2 \co(u_{13}+u_{23}+\zeta_{1}-\zeta_{2})+2 \co(u_{13}+u_{23}-\zeta_{1}+\zeta_{2})\\
&-4 \co(2 \zeta_{3}) \s(u_{12}) \s(\zeta_{1}+\zeta_{2})\Bigr],\\
Y^{111}_{011}&=\frac{2\mi}{\mathcal{N}}\Bigl[\left\{\co(u_{23})+\co(\zeta_{2}+\zeta_{3})\right\} \left\{ \s(u_{13})\s(\zeta_{1}+\zeta_{2})+\s(u_{12})\s(\zeta_{1}-\zeta_{3})\right\}\Bigl],\\
Y^{111}_{110}&=\frac{2\mi}{\mathcal{N}}\Bigl[\left\{\co(u_{12})+\co(\zeta_{1}+\zeta_{2})\right\} \left\{ \s(u_{23})\s(\zeta_{1}-\zeta_{3})-\s(u_{13})\s(\zeta_{2}+\zeta_{3})\right\}\Bigl],\\
Y^{111}_{101}&=\frac{2\mi}{\mathcal{N}}\Bigl[\left\{\co(u_{13})+\co(\zeta_{1}-\zeta_{3})\right\}\left\{\s(u_{23})\s(\zeta_{1}+\zeta_{2})-\s(u_{12})\s(\zeta_{2}+\zeta_{3})\right\}\Bigr],\\
Y^{111}_{001}&=\frac{2}{\mathcal{N}}\Bigl[\left\{\co(u_{13}+\co(\zeta_{1}-\zeta_{3})\right\}\left\{\co(u_{23})+\co(\zeta_{2}+\zeta_{3})\right\}\s(\zeta_{1}+\zeta_{2})\Bigr],\\
          \displaybreak \\ 
Y^{111}_{010}&=\frac{2}{\mathcal{N}}\Bigl[\left\{\co(u_{12})+\co(\zeta_{1}+\zeta_{2})\right\}\left\{\co(u_{23})+\co(\zeta_{2}+\zeta_{3})\right\}\s(\zeta_{1}-\zeta_{3})\Bigr],\\
Y^{111}_{100}&=-\frac{2}{\mathcal{N}}\Bigl[\left\{\co(u_{12})+\co(\zeta_{1})+\zeta_{2})\right\} \left\{\co(u_{13})+\co(\zeta_{1}-\zeta_{3})\right\}\s(\zeta_{2}+\zeta_{3})\Bigr].
\end{align*}

In the above formulas, the factor  $\mathcal{N}$ is given by the following expression:
\begin{align}
    \mathcal{N}&=\left\{\co(u_{12}) - \co(u_{13}+u_{23})\right\}\s(\zeta_{1}+\zeta_{2})+2\s(u_{12})\s(u_{23})\s(\zeta_{1}-\zeta_{3})\label{app:normfactor}\\
&-2\s(u_{12})\s(u_{13})\s(\zeta_{2}+\zeta_{3})-2\s(\zeta_{1}+\zeta_{2})\s(\zeta_{1}-\zeta_{3})\s(\zeta_{2}+\zeta_{3})\nonumber.
\end{align}

The other coefficients are obtained from the ones above from the relations:\\
\begin{minipage}[t]{.3\textwidth}
%\raggedright
\begin{align*}
S^{111}_{001}&= S^{010}_{001}+Y^{111}_{001};\\
S^{111}_{011}&=S^{010}_{011}+Y^{111}_{011};\\
S^{111}_{100}&= S^{010}_{100}+Y^{111}_{100};\\
S^{111}_{101}&= S^{010}_{101}+Y^{111}_{101};\\
S^{111}_{110}&= S^{010}_{110}+Y^{111}_{110};\\
S^{111}_{010}&=S^{010}_{010}+Y^{111}_{010}-1;\\
\end{align*}
\end{minipage}% <---------------- Note the use of "%"
\begin{minipage}[t]{.3\textwidth}
%\raggedleft
\begin{align*}
S^{100}_{010}&= S^{001}_{010};\\
S^{100}_{001}&=S^{001}_{001}+1;\\
S^{100}_{011}&= S^{001}_{011};\\
S^{100}_{101}&= S^{001}_{101};\\
S^{100}_{110}&= S^{001}_{110};\\
S^{100}_{100}&= S^{001}_{100}-1;\\
\end{align*}
\end{minipage}
\begin{minipage}[t]{.3\textwidth}
\begin{align*}
S^{001}_{001}&= -S^{010}_{001};\\
S^{001}_{010}&= 1-S^{010}_{010};\\
S^{001}_{100}&= 1-S^{010}_{100};\\
S^{001}_{011}&= -S^{010}_{011};\\
S^{001}_{101}&= -S^{010}_{101};\\
S^{001}_{110}&= -S^{010}_{110}.
\end{align*}
\end{minipage}
%% Loading bibliography style file
\bibliographystyle{utphys}
\bibliography{zta_final}
\end{document}